# Thermodynamically Consistent Darcy-Brinkman-Forchheimer Framework in Matrix Acidization*

Yuanqing Wu[1], Jisheng Kou[2#], Shuyu Sun[3#], Yu-Shu Wu[4]

[1]*College of Mathematics and Statistics, Shenzhen University, Shenzhen, Guangdong 518048, China*

[2]*School of Civil Engineering, Shaoxing University, Shaoxing, Zhejiang 312000, China*

[3]*Computational Transport Phenomena Laboratory, Division of Physical Science and Engineering, King Abdullah University of Science and Technology, Thuwal 23955-6900, Saudi Arabia*

[4]*Department of Petroleum Engineering, Colorado School of Mines, 1600 Arapahoe Street, Goden, CO 80401, USA*

## Abstract

Matrix acidization is an important technique to enhance oil production at the tertiary recovery stage, and its numerical simulation is never concluded. From one of the earliest models, i.e. the two-scale model (Darcy framework), the Darcy-Brinkman-Forchheimer (DBF) framework is developed by adding Brinkman term and Forchheimer term to the momentum conservation equation. However, in the momentum conservation equation of the DBF framework, porosity is put outside of the time derivation term, which cannot describe the change of porosity well. Thus, this work changes the expression so that the modified momentum conservation equation can satisfy Newton's second law. The modified framework is called improved DBF framework. Furthermore, based on the improved DBF framework, the thermal DBF framework is given by introducing the energy balance equation to the improved DBF framework. Both of the frameworks are verified by the former works through numerical experiments and chemical experiments in labs. Parallelization to the codes of the complicated frameworks is also realized, and good scalability can be achieved.

## 1. Introduction

Acidization is a useful technique to promote or restore oil production in reservoirs, which can be classified into two types: fracture acidization and matrix acidization. In fracture acidization, a highly-pressurized acid flow is injected into the well to physically enlarge the fractures and chemically dissolve the permeability inhibitive deposits. However, in matrix acidization, the pressure of the acid flow is not high enough to destroy the fractures, and thus the acid flow can only enlarge the natural pores of the matrix. In a word, both kinds of acidization try to enlarge the voids of reservoirs and ease the outflow of hydrocarbons from subsurface matrix. A lot of works [1]- [3] have studied fracture acidization, but this work pays attention on matrix acidization.

Theoretically, matrix acidization is one topic of the chemical dissolution-front instability problem, and many studies focus on the factors affecting matrix acidization such as the mineral reactive surface area [4], mineral dissolution ratio [5], solute dispersion [6], etc. Numerically, four main models have been proposed to investigate matrix acidization, including the capillary tube model [7], the network model [8]-[10], the dimensionless model [11]-[12] and the two-scale model [13]-[15]. Since the two-scale model can predict the dissolution patterns better and capture the formation of wormholes more

* This work is supported by Peacock Plan Foundation of Shenzhen (No. 000255), National Natural Science Foundation of China (No. 11601345) and Natural Science Foundation of SZU (No. 2017059).

# Corresponding Author. E-mail address: koujisheng@163.com, shuyu.sun@kaust.edu.sa.

accurately, this work focuses on it. In fact, the two-scale means the Darcy scale and pore scale. In each scale, there are a series of equations to describe the progress of matrix acidization. At the initial stage of the model, the momentum conservation equation in the Darcy scale is written from Darcy's law, so the two-scale model can also be called the Darcy framework in this work [15]- [18]. Later, due to the nature of matrix acidization, Wu et al. [19]-[22] consider the Darcy framework is not accurate enough to describe matrix acidization, and provide the Darcy-Brinkman-Forchheimer (DBF) framework by adding Brinkman term and Forchheimer term to the momentum conservation equation. Li et al. [23]-[25] further analyse the numerical stability and accuracy of the DBF framework. However, the DBF framework still has a defect when processing the momentum conservation equation, which degrades its reliability. This work reviews the DBF framework in more details than [19], and points out the defect, which will be shown in Section 2. Moreover, in the DBF framework, a pseudo parameter $\varepsilon$ is introduced into the mass conservation equation to solve the linear system by an iterative solver HYPRE [26], which also degrades the reliability of the DBF framework. However, by replacing HYPRE with a direct solver MUMPS [27][28], the introduction of $\varepsilon$ is not necessary, which is realized in this work. Furthermore, the flowchart of simulation should be also changed accordingly. After these, the new framework provided in this work is called the improved two-scale model based on the DBF framework, or improved DBF framework for short.

However, all the frameworks above only consider the mass conservation law and momentum conservation law, and another kind of conservation that is the energy conservation law is not included, which is a main drawback of them. As we know, the temperature is a key variable in the energy conservation law, and has a significant influence on the thermodynamic parameters such as the surface reaction rate and molecular diffusion coefficient. However, these thermodynamic parameters are deemed as constants in those frameworks. Moreover, in real applications, the operation of matrix acidization is done in subsurface environment where matrix is warmed by terrestrial heat. Thus, the temperature should have been considered as a main factor in matrix acidization, which brings about the necessity and reasonability of introducing the energy conservation equation to the frameworks.

Besides this work, a lot of works have noticed the temperature issue, and upgraded the two-scale model by their own ways. For example, Li et al. [29] introduce a heat transmission equation in the form of radial flows to the two-scale model, and gives out the simulation results near wellbore. Ma et al. [2] also develop a temperature-influenced model based on the two-scale model, and use it to simulate matrix acidization in fractured carbonate rocks. Kalia et al. [30] study the cases when the temperatures of the acid fluid and the matrix are different, and simulate acidization in the matrix of both adiabatic and non-adiabatic conditions. They conclude that the fluid temperature can be designed as a parameter to control matrix acidization. Although these endeavors try to involve the thermal effect on matrix acidization into consideration, all of them are based on the Darcy framework that is not accurate enough to simulate matrix acidization as mentioned above. As a result, their reliability is degraded. Therefore, this work provides a heat transfer model as an expansion of the more reasonable improved DBF framework, and aims to output more reliable results. The new model is the thermodynamically consistent DBF framework, which can be called the thermal DBF framework for short. Different from [29], this work studies matrix acidization in the form of linear flows. Meanwhile, fractures in the matrix are not considered, and a general matrix is acidized, which is different from [2]. Inspired from [30], the two cases when the temperatures of the acid flow and the matrix are the same and different are studied, respectively, and the results are verified against [31] and [30], respectively.

The work [19] only realizes the 2D parallel code of the DBF framework, and this work further realizes the 3D parallel code of the DBF framework. Not only that, the 2D and 3D parallel codes of the improved DBF framework and thermal DBF framework have also been realized in this work.

In the following discussions, the improved DBF framework is developed firstly, and then based on it, the thermal DBF framework is provided. In the model verification section, the correctness of the improved DBF framework is checked, by the way of comparing its 2D and 3D results with the existing works. Only when the correctness of the improved DBF framework is guaranteed can the reasonability of the thermal DBF framework be assured. After this, a series of thermal experiments are carried out to investigate the temperature effect on matrix acidization. The performance of the 3D parallel code is evaluated at the end of this work.

## 2. Improved DBF Framework and Its Solution Scheme

In the statements below, the meanings of all the notations are given in Table 1. The Darcy framework is used very popularly to simulate the matrix acidization procedure. In the pore scale, a group of semi-empirical equations is provided to describe the relationship of parameters in the pore scale such as porosity, permeability, and local mass-transfer coefficient. These equations have few changes during the study of matrix acidization simulation. However, in the Darcy scale, which is the other scale of the Darcy framework, the equations have a lot of changes with the progress of the study. Generally speaking, there are three kinds of equations in the Darcy scale: momentum conservation equations, mass conservation equations, and concentration balance equations. The changes mainly lie in the momentum conservation equations, while the other two kinds of equations keep more or less the same, although some small changes are made to them according to the need of different cases. Initially, the momentum conservation equation is represented by Darcy's law

$$\nabla p + \frac{\mu}{K}\boldsymbol{u} = 0,$$

supposing the permeability is homogeneous and isotropic. However, the applicability of this equation is limited to the condition where Reynolds number $Re < 1$ and Darcy number $Da \ll 1$. At the beginning of matrix acidization, the porosity in the porous medium is not very large, and Darcy's law can be leveraged to describe the flow of fluid in the porous medium properly. However, with the propagation of the channels due to matrix acidization, the areas eaten by the channels become large. In the channels, porosity can be very large, and even approach the value one, which brings about high permeability in these areas. From the definition of Darcy number, it is learned that Darcy number is much higher than one as a result of that. Moreover, in the high-permeability areas, the velocity of fluid becomes large, which may lead to a high Reynolds number that can be far from the value "one" assuming that the viscosity and mass density of the fluid keep more or less the same and the particle size is constant. All of these are the reasons that Darcy's law is not suitable to be used in matrix acidization simulation.

In order to fix the issue, some corrections have to be made to Darcy's law to cope with the conditions of high permeability and high Reynolds numbers. The first correction is called Brinkman correction. According to Darcy's law, the uniform velocity in the cross-sectional direction can be seen when the permeability is low. However, in the porous medium of high permeability, the no-slip condition should be considered instead of uniform velocity. Brinkman correction introduces a viscous shear stress term to Darcy's law, by which the no-slip condition can be described well. The Brinkman-corrected Darcy's law can be expressed as

$$\nabla p + \frac{\mu}{K} \boldsymbol{u} - \frac{\mu}{\phi} \nabla^2 \boldsymbol{u} = 0,$$

in which $\frac{\mu}{\phi} \nabla^2 \boldsymbol{u}$ is called Brinkman term. Besides that, in the condition of high Reynolds numbers, form drag can be much larger than viscous drag, which can be suitably described by Forchheimer correction. In this correction, a term called Forchheimer term, which is expressed as $\frac{\rho_f F}{\sqrt{K}} |\boldsymbol{u}| \boldsymbol{u}$, is added to Darcy's law. Combining both of the two corrections together, Darcy's law can be modified as

$$\nabla p + \frac{\mu}{K} \boldsymbol{u} - \frac{\mu}{\phi} \nabla^2 \boldsymbol{u} + \frac{\rho_f F}{\sqrt{K}} |\boldsymbol{u}| \boldsymbol{u} = 0,$$

with $F = \frac{1.75}{\sqrt{150 \phi^3}}$ as Forchheimer coefficient [32]. In the equation above, the right-hand side equals with zero, which means that the sum of all the external forces imposed on the fluid is zero. However, it is not a general case. For the cases that the sum of all the external forces is not zero, the right-hand side of the equation should equal with the product of mass density and acceleration. With the Eulerian expression of acceleration, the right-hand side of the equation can be written as

$$-\frac{\rho_f}{\phi} \frac{\partial \boldsymbol{u}}{\partial t} - \frac{\rho_f}{\phi^2} \nabla \cdot \boldsymbol{u} \boldsymbol{u}.$$

If the gravity effect is considered, the momentum conservation equation in its final form can be written as

$$\frac{\rho_f}{\phi} \frac{\partial \boldsymbol{u}}{\partial t} + \frac{\rho_f}{\phi^2} \nabla \cdot \boldsymbol{u} \boldsymbol{u} = -\nabla p - \frac{\mu}{K} \boldsymbol{u} + \frac{\mu}{\phi} \nabla^2 \boldsymbol{u} - \frac{\rho_f F}{\sqrt{K}} |\boldsymbol{u}| \boldsymbol{u} + \rho_f \boldsymbol{g}. \qquad (1)$$

Equation (1) is the momentum conservation equation used in our former work [19]. Because such kind of momentum conservation equation introduces Brinkman term and Forchheimer term, this model is called the two-scale model based on the Darcy-Brinkman-Forchheimer (DBF) framework, or the DBF framework for short. More details of it can be referenced to [19]. However, the framework cannot meet Newton's second law, since porosity $\phi$ is changed during the simulation procedure. It is noted that in the first term on the left-hand side of Equation (1), $\phi$ is outside of the time derivative, which hints that $\phi$ does not change with time. This contradicts the true physical observation, as a result of which Newton's second law is also violated. Thus, $\phi$ should be put inside the time derivative. After this operation, $\frac{\boldsymbol{u}}{\phi}$ can be deemed as a new variable, which is the effective velocity. Accordingly, the second term on the left-hand side of Equation (1) and the third term on the right-hand side of Equation (1) are changed, which can be expressed as

$$\rho_f \frac{\partial}{\partial t} \left( \frac{\boldsymbol{u}}{\phi} \right) + \rho_f \frac{\boldsymbol{u}}{\phi} \cdot \nabla \frac{\boldsymbol{u}}{\phi} = -\nabla p - \frac{\mu}{K} \boldsymbol{u} + \nabla \cdot \mu \nabla \frac{\boldsymbol{u}}{\phi} - \frac{\rho_f F}{\sqrt{K}} |\boldsymbol{u}| \boldsymbol{u} + \rho_f \boldsymbol{g}. \qquad (2)$$

It is noted that the left-hand side of Equation (2) is in reality the material derivative

$$\rho_f \frac{\partial}{\partial t} \left( \frac{\boldsymbol{u}}{\phi} \right) + \rho_f \frac{\boldsymbol{u}}{\phi} \cdot \nabla \frac{\boldsymbol{u}}{\phi} = \rho_f \frac{D}{Dt} \left( \frac{\boldsymbol{u}}{\phi} \right).$$

The discussions above demonstrate from the theories of fluid dynamics that the new momentum conservation equation should describe matrix acidization more reasonably, and the new model is called the "improved" two-scale model based on the DBF framework, or the improved DBF framework for short.

Since the flow in matrix acidization is supposed to be incompressible, the mass conservation equation can be expressed as

$$\nabla \cdot \boldsymbol{u} = 0.$$

However, considering the local volume change in the matrix acidization procedure, the mass conservation equation should be modified as

$$\frac{\partial \phi}{\partial t} + \nabla \cdot \boldsymbol{u} = 0. \qquad (3)$$

The concentration balance equation can be derived from the principle of species balance during matrix acidization. The balance of species can be achieved by accumulation, advection, diffusion, and reaction effects, which brings about the straightforward expression of the concentration balance equation

$$\frac{\partial (\phi C_f)}{\partial t} + \nabla \cdot \left( \boldsymbol{u} C_f \right) = \nabla \cdot \left( \phi \boldsymbol{D_e} \cdot \nabla C_f \right) - k_c a_v \left( C_f - C_s \right). \qquad (4)$$

In the equation, $\boldsymbol{D_e}$ is a function of $\boldsymbol{u}$

$$\boldsymbol{D_e} = d_m \boldsymbol{I} + \|\boldsymbol{u}\|(d_l \boldsymbol{E} + d_t \boldsymbol{E^\perp}),$$

with

$$d_l = \alpha_{OS} d_m + \frac{2\lambda_X \|\boldsymbol{u}\| \bar{r}_p}{\phi},$$

$$d_t = \alpha_{OS} d_m + \frac{2\lambda_T \|\boldsymbol{u}\| \bar{r}_p}{\phi}.$$

In the 3D condition,

$$\boldsymbol{E} = \frac{1}{\|\boldsymbol{u}\|^2} \begin{pmatrix} u_x^2 & u_x u_y & u_x u_z \\ u_y u_x & u_y^2 & u_y u_z \\ u_z u_x & u_z u_y & u_z^2 \end{pmatrix},$$

$$\boldsymbol{E^\perp} = \boldsymbol{I} - \boldsymbol{E}.$$

$u_x$, $u_y$, and $u_z$ stand for the $x$-direction, $y$-direction, and $z$-direction velocity, respectively. The 2D condition is similar. In fact, the concentration balance equation is the other expression of the mass conservation law, besides the mass conservation equation.

In Equation (2), (3), and (4), the velocity vector $\boldsymbol{u}$, pressure $p$, and the cup-mixing concentration of the acid $C_f$ are deemed as unknowns to be solved for. However, the three equations are not enough to do that, since the values of the other variables are unknown. Thus, more auxiliary equations and some necessary assumptions are needed. Three additional equations are given as below

$$k_c(C_f - C_s) = R(C_s, T), \qquad (5)$$

$$\frac{\partial \phi}{\partial t} = \frac{R(C_s)a_v\alpha}{\rho_s}, \qquad (6)$$

$$C_s = \frac{C_f}{1 + \frac{k_s}{k_c}}, \qquad (7)$$

These equations are derived from mathematical deduction and chemical experiments, and more details can be referred to [15]. It is emphasized that Equation (7) can be put in Equation (4) to substitute $C_s$ when solving $C_f$. Moreover, since the mass density $\rho_f$ and viscosity $\mu$ of the fluid will not change much during the matrix acidization procedure, they are supposed to be given constants for simplicity. It is also easy to understand that the dissolving power of the acid $\alpha$ and the mass density of the solid phase $\rho_s$ can also be deemed as given constants. For the reason mentioned later, the surface reaction rate $k_s$ is no longer deemed as a constant as shown in [19]. Instead, it is a function of the temperature $T$ in this work. As a result, the reaction rate also becomes a function of $T$, and can be rewritten as $R(C_s, T)$, which is different from [19].

Besides that, it is noted that some variables in the pore scale such as porosity, permeability, the interfacial surface area per unit volume and the local mass-transfer coefficient appear in the equations of Darcy scale. Thus, their values should be known before we try to solve for the unknowns in the Darcy scale, with the help of a series of equations in the pore scale. It is stipulated that the subscript 0 represents the initial value or reference value of the corresponding variable in the following equations, and all the initial values are known. Firstly, three equations called Carman-Kozeny correlation are provided as below

$$\frac{K}{K_0} = \frac{\phi}{\phi_0}\left(\frac{\phi(1-\phi_0)}{\phi_0(1-\phi)}\right)^2, \qquad (8)$$

$$\frac{r_p}{r_0} = \sqrt{\frac{K\phi_0}{K_0\phi}}, \qquad (9)$$

$$\frac{a_v}{a_0} = \frac{\phi r_0}{\phi_0 r_p}. \qquad (10)$$

From these equations, it is learned how permeability, the pore radius, and the interfacial surface area per unit volume change with porosity. Thus, as long as the porosity is known, the values of the three variables can be computed from Carman-Kozeny correlation. Now we come to see how to compute porosity. From Equation (5), (6), and (7), the equation below can be derived

$$\frac{\partial \phi}{\partial t} = \frac{a_v\alpha C_f k_c k_s}{\rho_s(k_c + k_s)}. \qquad (11)$$

The left-hand side of Equation (11) describes the change of porosity with time, and its right-hand side includes a lot of variables among which, except $a_v$ and $k_c$, all the other variables have no direct relationship with porosity. By using Equation (8), (9), and (10), $a_v$ can be expressed as a function of $\phi$

$$a_v = a_0 \frac{1-\phi}{1-\phi_0}. \tag{12}$$

Equation (11) can then be changed as

$$\frac{\partial \phi}{\partial t} = \frac{a_0 \alpha C_f k_c k_s (1-\phi)}{\rho_s (k_c + k_s)(1-\phi_0)}. \tag{13}$$

Moreover, the local mass-transfer coefficient $k_c$ can be calculated from the expression of the Sherwood number $Sh$ which is a dimensionless mass-transfer coefficient. The expression is given as below

$$Sh = \frac{2k_c r_p}{d_m} = Sh_\infty + 0.7 Re^{1/2} Sc^{1/3}. \tag{14}$$

On the right-hand side of Equation (14), $Sh_\infty$ is a given constant, and the Reynolds number $Re$ can be expressed as

$$Re = \frac{2\|\boldsymbol{u}\| r_p \rho_f}{\mu}. \tag{15}$$

The Schmidt number $Sc$ is expressed as

$$Sc = \frac{\mu d_m}{\rho_f}. \tag{16}$$

Thus, by leveraging Equation (14), (15), and (16) together, $k_c$ can be calculated as

$$k_c = \frac{d_m}{2r_p} \left( Sh_\infty + 0.7 \left( \frac{2\|\boldsymbol{u}\| r_p \rho_f}{\mu} \right)^{\frac{1}{2}} \left( \frac{\mu d_m}{\rho_f} \right)^{\frac{1}{3}} \right). \tag{17}$$

It is observed that Equation (17) includes the variable $r_p$ that is a function of porosity from Equation (8) and (9),

$$r_p = r_0 \frac{\phi(1-\phi_0)}{\phi_0(1-\phi)}, \tag{18}$$

which means that $k_c$ is in fact a function of porosity and the velocity. Thus, Equation (17) can substitute $k_c$ in Equation (13), and then a new equation with porosity being the unknown can be derived. However, the new equation is too complex to derive the analytic formula of porosity, and therefore the value of porosity has to be computed by the numerical scheme. In that condition, the semi-implicit scheme is applied. In the following statements, the superscripts of the notations represent the time step. The porosity at the time step $\tau$ is used in Equation (18) to calculate the pore radius at the time step $\tau$ which is then put into Equation (17) to compute $k_c$ at the time step $\tau$. With the semi-implicit scheme, Equation (13) can be rewritten as

$$\frac{\phi^{\tau+1}-\phi^{\tau}}{\Delta t}=\frac{a_0\alpha C_f^{\tau}k_c^{\tau}k_s^{\tau}(1-\phi^{\tau+1})}{\rho_s(k_c^{\tau}+k_s^{\tau})(1-\phi_0)}, \tag{19}$$

by which $\phi^{\tau+1}$ can be computed easily.

From the discussions above, it can be learned that the main unknown to be solved for in the pore scale is porosity $\phi$. As long as the value of porosity is gotten, the values of the other variables in the pore scale can be derived from a series of pore-scale equations. Furthermore, with the values of all the pore-scale variables, the main unknowns in the Darcy scale can be calculated by the Darcy-scale equations. At the same time, from the computation of porosity, it is found that the variables $\boldsymbol{u}$ and $C_f$ in the Darcy scale will affect the value of porosity. Therefore, the computations of the Darcy scale and pore scale are coupled with each other, with the Darcy-scale variables $\boldsymbol{u}$ and $C_f$ and the pore-scale variable $\phi$ being their interaction media, which can be shown in Figure 1.

The derivation of all the equations in both of the Darcy scale and pore scale has been done in the former discussions, and then numerical schemes are applied on them to solve for the variables. Since porosity $\phi$ plays a central role in the improved DBF framework, it should be computed from Equation (19) with the semi-implicit scheme firstly. Then, with the computed porosity, permeability $K$, and the interfacial surface area per unit volume $a_v$ can be computed with Equation (8) and (12), respectively. Next, Equation (2) and (3) are combined together as a linear system and solved for the velocity $\boldsymbol{u}$ and pressure $p$ with the semi-implicit scheme. It is emphasized that the term $\frac{\partial \phi}{\partial t}$ in Equation (3) is substituted by Equation (6) when the linear system is solved. Moreover, with the update of porosity $\phi$, the local mass-transfer coefficient is also updated by Equation (17) to $k_c^{\tau+\frac{1}{2}}$ before solving the linear system. With the semi-implicit scheme, Equation (2) and (3) can be rewritten as below, respectively.

$$\rho_f\frac{\frac{\boldsymbol{u}^{\tau+1}}{\phi^{\tau+1}}-\frac{\boldsymbol{u}^{\tau}}{\phi^{\tau}}}{\Delta t}+\rho_f\frac{\boldsymbol{u}^{\tau}}{\phi^{\tau}}\cdot\nabla\frac{\boldsymbol{u}^{\tau+1}}{\phi^{\tau+1}}$$

$$=-\nabla p^{\tau+1}-\frac{\mu}{K^{\tau+1}}\boldsymbol{u}^{\tau+1}+\nabla\cdot\mu\nabla\frac{\boldsymbol{u}^{\tau+1}}{\phi^{\tau+1}}-\frac{\rho_f F^{\tau+1}}{\sqrt{K^{\tau+1}}}|\boldsymbol{u}^{\tau}|\boldsymbol{u}^{\tau+1}+\rho_f\boldsymbol{g},$$

$$\frac{a_v^{\tau+1}\alpha C_f^{\tau}k_c^{\tau+\frac{1}{2}}k_s^{\tau}}{\rho_s(k_c^{\tau+\frac{1}{2}}+k_s^{\tau})}+\nabla\cdot\boldsymbol{u}^{\tau+1}=0.$$

After that, since the velocity $\boldsymbol{u}$ are updated, the local mass-transfer coefficient can be updated again by Equation (17) to $k_c^{\tau+1}$. Lastly, the semi-implicit scheme is used to solve Equation (4) for concentration $C_f$, and then another linear system is formed. It is emphasized that Equation (7) is put into Equation (4) when the linear system is solved, which is shown as

$$\frac{\phi^{\tau+1}C_f^{\tau+1}-\phi^{\tau}C_f^{\tau}}{\Delta t}+\nabla\cdot\left(\boldsymbol{u}^{\tau+1}C_f^{\tau+1}\right)$$

$$= \nabla \cdot \left( \phi^{\tau+1} \boldsymbol{D}_e^{\tau+1} \cdot \nabla C_f^{\tau+1} \right) - k_c^{\tau+1} a_v^{\tau+1} \left( C_f^{\tau+1} - \frac{C_f^{\tau+1}}{1 + \frac{k_s^{\tau+1}}{k_c^{\tau+1}}} \right). \qquad (20)$$

In brief, the solution procedure can be described as a flowchart, which is shown in Figure 2.

From the flowchart, another difference of this work from our former work [19] can be seen. In the former work, for each iteration, the simulation begins with the computation of the variables in the Darcy scale such as pressure, velocity, and concentration, and ends with the computation of the variables in the pore scale such as porosity, permeability, and the interfacial surface area per unit volume. However, in this work, the computation of the variables in the pore scale is ahead of the computation of the variables in the Darcy scale. The flowchart of this work is more reasonable, which can be demonstrated by the following statements. From Equation (20), it is observed that in order to get $C_f^{\tau+1}$, we have to know $\phi^{\tau+1}$, and therefore the computation of $\phi^{\tau+1}$ should be ahead of the computation of $C_f^{\tau+1}$. However, in the former work, the computation of $\phi^{\tau+1}$ is behind the computation of $C_f^{\tau+1}$, which is not reasonable. With the philosophy, in the simulation of the improved DBF framework, it is suggested to compute the variables in the pore scale at first, followed by the variables of the Darcy scale.

### 3. Thermal DBF Framework and Its Solution Scheme

Based on the improved DBF framework, a heat transfer model which considers the heat transmission process in matrix acidization is developed. The model is composed of the improved DBF framework and the energy conservation equation which can be expressed as the governing equation of the temperature $T$

$$\frac{\partial \vartheta}{\partial t} + \nabla \cdot \left( \boldsymbol{u} \frac{\vartheta_f}{\phi} \right)$$

$$= \nabla \cdot \lambda \nabla T - p \nabla \cdot \boldsymbol{u} + \mu \nabla \frac{\boldsymbol{u}}{\phi} : \nabla \frac{\boldsymbol{u}}{\phi} + \frac{\mu}{K} |\boldsymbol{u}|^2 + \frac{\rho_f F}{\sqrt{K}} |\boldsymbol{u}|^3 + a_v R(C_s, T) H_r(T), \qquad (21)$$

where

$$\vartheta = \vartheta_f + \vartheta_s,$$

$$\vartheta_f = \phi \rho_f \theta_f T,$$

$$\vartheta_s = (1 - \phi) \rho_s \theta_s T,$$

$$\lambda = \phi \lambda_f + (1 - \phi) \lambda_s,$$

$$H_r(T) = |-9702 + 16.97T - 0.00234T^2|.$$

$H_r(T)$ is the reaction heat [29]. $\lambda_f$ and $\lambda_s$ are the heat conduction coefficients of the fluid and solid phase, respectively, and thus $\lambda$ is the average heat conduction coefficient between the two phases. $\theta_f$ and $\theta_s$ are the heat capacities of the fluid and solid phase, respectively. $\lambda_f$, $\lambda_s$, $\theta_f$, and $\theta_s$ are deemed as constants in this work. $\vartheta_f$ and $\vartheta_s$ are the amounts of heat per unit volume of the fluid and solid phase, respectively, and thus $\vartheta$ is the total amount of heat per unit volume. The total energy is a sum of the fluid energy and solid medium energy, which may vary with time. The energy transportation caused by the fluid flow occurs only between fluids in different spatial positions due to fluid flow. Both the heat conductions in the interior of fluids and solids and with each other are considered in the first term of the right-hand side of (21). The works done by the pressure and viscosity force of the fluid are described in the second and the third terms of the right-hand side of (21). The works done by friction forces between fluids and solids are described in the fourth and the fifth terms of the right-hand side of (21) based on the Darcy-Forchheimer framework. The chemical reaction may produce the heat, which is considered in the last term of (21). All the terms constitute the source or sink of energy. It is emphasized that the temperature of the acid and matrix is assumed to become the same immediately when acid is injected into the matrix, since the speed of heat transfer is much faster than the fluid speed. Thus, the temperature of the acid and matrix can be represented by a single notation $T$. In fact, differentiating between the acid temperature and matrix temperature brings challenges to theories and applications, and the details of the heat transfer between the acid and matrix must be studied thoroughly. Relevant work can be left to the future. Furthermore, the surface reaction rate $k_s$ is deemed as a function of the temperature $T$, which can be expressed as

$$k_s = k_{s0} \cdot e^{\frac{E_g}{R_g}(\frac{1}{T_0} - \frac{1}{T})},\tag{22}$$

in which $k_{s0}$ is the surface reaction rate at temperature $T_0$, $E_g$ is the activation energy, and $R_g$ is the molar gas constant [30]. As a result of that, the reaction rate also becomes a function of $T$, and its expression is rewritten as $R(C_s, T)$ in (21). Moreover, the molecular diffusion coefficient $d_m$ is also affected by the temperature $T$, which can be expressed as [29]

$$d_m = d_{m0} \cdot e^{\frac{E_g}{R_g}(\frac{1}{T_0} - \frac{1}{T})}.\tag{23}$$

$d_{m0}$ is the molecular diffusion coefficient at temperature $T_0$. The heat transfer model induced from the improved DBF framework can be called the thermal DBF framework for short.

After solving a series of equations in the improved DBF framework, the semi-implicit scheme can be used to solve Equation (21) for the temperature $T$, which can be expressed as

$$\frac{(\phi^{\tau+1}\rho_f\theta_f T^{\tau+1} + (1-\phi^{\tau+1})\rho_s\theta_s T^{\tau+1}) - (\phi^{\tau}\rho_f\theta_f T^{\tau} + (1-\phi^{\tau})\rho_s\theta_s T^{\tau})}{\Delta t} + \nabla \cdot \left( \boldsymbol{u}^{\tau+1}\phi^{\tau+1}\rho_f\theta_f T^{\tau+1} \right)$$

$$= \nabla \cdot \left( \phi^{\tau+1}\lambda_f + (1-\phi^{\tau+1})\lambda_s \right) \nabla T^{\tau+1} - p^{\tau+1}\nabla \cdot \boldsymbol{u}^{\tau+1} + \mu \nabla \frac{\boldsymbol{u}^{\tau+1}}{\phi^{\tau+1}} : \nabla \frac{\boldsymbol{u}^{\tau+1}}{\phi^{\tau+1}}$$

$$+ \frac{\mu}{K^{\tau+1}} |\boldsymbol{u}^{\tau+1}|^2 + \frac{\rho_f F^{\tau+1}}{\sqrt{K^{\tau+1}}} |\boldsymbol{u}^{\tau+1}|^3 + k_c^{\tau+1} a_v^{\tau+1} \left( C_f^{\tau+1} - \frac{C_f^{\tau+1}}{1 + \frac{k_s^\tau}{k_c^{\tau+1}}} \right) H_r(T^\tau),$$

and the third linear system is formed. It is noted that Equation (7) is put into (21) in the above expression. When using the thermal DBF framework to simulate matrix acidization, the flowchart can be shown in Figure 3. From the figure, it can be seen that the molecular diffusion coefficient $d_m$ and the surface reaction rate $k_s$ are calculated at first, followed by a series of computations in the improved DBF framework, and the computation of the temperature is done lastly. It is also learnt that by the variables $\phi$, $\boldsymbol{u}$, $p$, and $C_f$, the improved DBF framework changes the main variable $T$ in the energy conservation equation, while the energy conservation equation also changes the variables of the improved DBF framework by $T$, or $k_s$ and $d_m$, which can be shown in Figure 4. From these discussions, it is known that the thermal DBF framework considers all the three kinds of conservation laws: mass, momentum, and energy, and should simulate matrix acidization more reasonably.

At the end of this section, the relationship among the Darcy framework, the DBF framework, the improved DBF framework and the thermal DBF framework can be summarized. Initially, the Darcy framework is provided to simulate matrix acidization, and achieves great success. In the framework, the communication between the pore scale and Darcy scale pushed forward the progress of simulation. Then, considering the clear fluid area which cannot be described accurately by Darcy's law, Brinkman term and Forchheimer term are introduced to the momentum conservation equation, and the DBF framework is developed. However, this framework cannot obey Newton's second law, and a modification is done, as a result of which the improved DBF framework is suggested. Until now, all the frameworks only consider two kinds of conservations, i.e. the mass conservation and the momentum conservation, which is not enough. Thus, based on the improved DBF framework, the third kind of conservation which is the energy conservation is introduced in the thermal DBF framework. Their relationship can also be seen in Figure 5.

## 4. Discretization and Parallelization

Possible discretization methods include the multipoint flux approximation method, the hybrid finite volume method, etc. The multipoint flux approximation method is "designed to give a correct discretization of the flow equations for general nonorthogonal grids as well as for general orientation of the principal directions of the permeability tensor" [33], while the hybrid finite volume method is "the ideal method for computing discontinuous solutions arising in compressible flows" [34]. Since this work considers orthogonal grids and incompressible flows, the finite difference method should be better to be chosen from all the possible discretization methods. In the following discussions, the finite difference method is used in discretizing the model, and the experimenting field approach [35]- [39] is used to compute the coefficients in the two linear systems. Although they have been used in the 2D simulation of matrix acidization in the former work [19], it is necessary to expand them to the 3D simulation which is one of the focuses of this work.

The equations used in the thermal DBF framework are discretized one by one according to the flowchart shown in Figure 3. For simplicity, suppose there is a 3D Cartesian grid. For Equation (22), (19), (8), (12), and (23), every variable is imposed at the centre of the cube. For Equation (17), except the variable $\boldsymbol{u}$, the other variables are imposed at the centre of the cube. However, $\|\boldsymbol{u}\|$ should also be imposed at the centre of the cube. Generally speaking, $\boldsymbol{u} = (u_x, u_y, u_z)$ and its three components are imposed on the faces of the cube, respectively. In other words, the $x$-direction velocity $u_x$ is imposed on the $x$-direction face which is vertical to the $x$-axis. The process of the $y$-direction velocity $u_y$ and $z$-direction velocity $u_z$ is similar. Thus, for a cube with its $x$-coordinate being from $i$ to $i+1$, $y$-coordinate being from $j$ to $j+1$ and $z$-coordinate being from $k$ to $k+1$, there is

$$\|\boldsymbol{u}\|_{\left(i+\frac{1}{2}, j+\frac{1}{2}, k+\frac{1}{2}\right)}$$

$$= \sqrt{(u_{x,i+1,j+\frac{1}{2},k+\frac{1}{2}} - u_{x,i,j+\frac{1}{2},k+\frac{1}{2}})^2 + (u_{y,i+\frac{1}{2},j+1,k+\frac{1}{2}} - u_{y,i+\frac{1}{2},j,k+\frac{1}{2}})^2 + (u_{z,i+\frac{1}{2},j+\frac{1}{2},k+1} - u_{z,i+\frac{1}{2},j+\frac{1}{2},k})^2},$$

with the subscript representing the coordinate. For Equation (2), it is in fact discretized on the faces of the cube, with the $x$-direction momentum equation being discretized on the $x$-direction face, the $y$-direction momentum equation being discretized on the $y$-direction face and the $z$-direction momentum equation being discretized on the $z$-direction face. Thus, the porosity and permeability on the faces should be known. However, from the discussions above, it is learnt that the porosity and permeability are imposed at the center of the cube. Thus, the harmonic method has to be applied to get their values on the faces. It is emphasized that the advection term in Equation (2) is discretized with the upwind scheme. After all the operations, the $x$-direction momentum equation imposed on the $x$-direction face with its $x$-coordinate being $i$, $y$-coordinate being from $j$ to $j+1$ and $z$-coordinate being from $k$ to $k+1$ can be discretized as below

$$\rho_f \frac{\dfrac{u^{\tau+1}_{x,i,j+\frac{1}{2},k+\frac{1}{2}}}{\phi^{\tau+1}_{i,j+\frac{1}{2},k+\frac{1}{2}}} - \dfrac{u^{\tau}_{x,i,j+\frac{1}{2},k+\frac{1}{2}}}{\phi^{\tau}_{i,j+\frac{1}{2},k+\frac{1}{2}}}}{\Delta t}$$

$$+ \frac{\rho_f}{\phi^{\tau}_{i,j+\frac{1}{2},k+\frac{1}{2}}} \left( \begin{array}{c} u^{\tau}_{x,i,j+\frac{1}{2},k+\frac{1}{2}} * \dfrac{\dfrac{u^{\tau+1}_{x,i,j+\frac{1}{2},k+\frac{1}{2}}}{\phi^{\tau+1}_{i,j+\frac{1}{2},k+\frac{1}{2}}} - \dfrac{u^{\tau+1}_{x,i-1,j+\frac{1}{2},k+\frac{1}{2}}}{\phi^{\tau+1}_{i-1,j+\frac{1}{2},k+\frac{1}{2}}}}{\Delta x} + \\[4ex] \bar{u}^{\tau}_{y,i,j+\frac{1}{2},k+\frac{1}{2}} * \dfrac{\dfrac{u^{\tau+1}_{x,i,j+\frac{1}{2},k+\frac{1}{2}}}{\phi^{\tau+1}_{i,j+\frac{1}{2},k+\frac{1}{2}}} - \dfrac{u^{\tau+1}_{x,i,j-\frac{1}{2},k+\frac{1}{2}}}{\phi^{\tau+1}_{i,j-\frac{1}{2},k+\frac{1}{2}}}}{\Delta y} + \bar{u}^{\tau}_{z,i,j+\frac{1}{2},k+\frac{1}{2}} * \dfrac{\dfrac{u^{\tau+1}_{x,i,j+\frac{1}{2},k+\frac{1}{2}}}{\phi^{\tau+1}_{i,j+\frac{1}{2},k+\frac{1}{2}}} - \dfrac{u^{\tau+1}_{x,i,j+\frac{1}{2},k-\frac{1}{2}}}{\phi^{\tau+1}_{i,j+\frac{1}{2},k-\frac{1}{2}}}}{\Delta z} \end{array} \right)$$

$$= - \frac{p^{\tau+1}_{i+\frac{1}{2},j+\frac{1}{2},k+\frac{1}{2}} - p^{\tau+1}_{i-\frac{1}{2},j+\frac{1}{2},k+\frac{1}{2}}}{\Delta x} - \frac{\mu}{K^{\tau+1}_{i,j+\frac{1}{2},k+\frac{1}{2}}} u^{\tau+1}_{x,i,j+\frac{1}{2},k+\frac{1}{2}}$$

$$
+\mu\left(\frac{\frac{\frac{u^{\tau+1}_{x,i+1,j+\frac{1}{2},k+\frac{1}{2}}}{\phi^{\tau+1}_{i+1,j+\frac{1}{2},k+\frac{1}{2}}}-\frac{u^{\tau+1}_{x,i,j+\frac{1}{2},k+\frac{1}{2}}}{\phi^{\tau+1}_{i,j+\frac{1}{2},k+\frac{1}{2}}}}{\Delta x}-\frac{\frac{u^{\tau+1}_{x,i,j+\frac{1}{2},k+\frac{1}{2}}}{\phi^{\tau+1}_{i,j+\frac{1}{2},k+\frac{1}{2}}}-\frac{u^{\tau+1}_{x,i-1,j+\frac{1}{2},k+\frac{1}{2}}}{\phi^{\tau+1}_{i-1,j+\frac{1}{2},k+\frac{1}{2}}}}{\Delta x}}{\Delta x}+\right.
$$
$$
\left.\frac{\frac{\frac{u^{\tau+1}_{x,i,j+\frac{3}{2},k+\frac{1}{2}}}{\phi^{\tau+1}_{i,j+\frac{3}{2},k+\frac{1}{2}}}-\frac{u^{\tau+1}_{x,i,j+\frac{1}{2},k+\frac{1}{2}}}{\phi^{\tau+1}_{i,j+\frac{1}{2},k+\frac{1}{2}}}}{\Delta y}-\frac{\frac{u^{\tau+1}_{x,i,j+\frac{1}{2},k+\frac{1}{2}}}{\phi^{\tau+1}_{i,j+\frac{1}{2},k+\frac{1}{2}}}-\frac{u^{\tau+1}_{x,i,j-\frac{1}{2},k+\frac{1}{2}}}{\phi^{\tau+1}_{i,j-\frac{1}{2},k+\frac{1}{2}}}}{\Delta y}}{\Delta y}+\frac{\frac{\frac{u^{\tau+1}_{x,i,j+\frac{1}{2},k+\frac{3}{2}}}{\phi^{\tau+1}_{i,j+\frac{1}{2},k+\frac{3}{2}}}-\frac{u^{\tau+1}_{x,i,j+\frac{1}{2},k+\frac{1}{2}}}{\phi^{\tau+1}_{i,j+\frac{1}{2},k+\frac{1}{2}}}}{\Delta z}-\frac{\frac{u^{\tau+1}_{x,i,j+\frac{1}{2},k+\frac{1}{2}}}{\phi^{\tau+1}_{i,j+\frac{1}{2},k+\frac{1}{2}}}-\frac{u^{\tau+1}_{x,i,j+\frac{1}{2},k-\frac{1}{2}}}{\phi^{\tau+1}_{i,j+\frac{1}{2},k-\frac{1}{2}}}}{\Delta z}}{\Delta z}\right)
$$

$$
-\frac{\rho_f F^{\tau+1}_{i,j+\frac{1}{2},k+\frac{1}{2}}}{\sqrt{K^{\tau+1}_{i,j+\frac{1}{2},k+\frac{1}{2}}}}\sqrt{\left(u^{\tau}_{x,i,j+\frac{1}{2},k+\frac{1}{2}}\right)^2+\left(\bar{u}^{\tau}_{y,i,j+\frac{1}{2},k+\frac{1}{2}}\right)^2+\left(\bar{u}^{\tau}_{z,i,j+\frac{1}{2},k+\frac{1}{2}}\right)^2}\,u^{\tau+1}_{x,i,j+\frac{1}{2},k+\frac{1}{2}}+\rho_f g_x,
$$

in which $\bar{u}_y$ and $\bar{u}_z$ represent the $y$-direction average velocity and $z$-direction average velocity on the face, respectively. $\bar{u}_y$ is computed as the average of the $y$-direction velocities on the four $y$-direction faces adjacent to the face. The computation of $\bar{u}_z$ is similar. $g_x$ is the $x$-direction component of $\boldsymbol{g}$. It is noted that in the above equation, it is assumed that

$$
u^{\tau}_{x,i,j+\frac{1}{2},k+\frac{1}{2}}>0,\ \bar{u}^{\tau}_{y,i,j+\frac{1}{2},k+\frac{1}{2}}>0\ \ and\ \ \bar{u}^{\tau}_{z,i,j+\frac{1}{2},k+\frac{1}{2}}>0.
$$

The discretization of the $y$-direction and $z$-direction momentum equation is similar. Under Neumann boundary condition for pressure, the momentum conservation equation on the boundary degenerates to

$$
u=u_B,
$$

in which $u_B$ is the boundary normal velocity. The discretization of Equation (3) is straightforward. For Equation (4), it is discretized at the centre of the cube. For a cube with its $x$-coordinate being from $i$ to $i+1$, $y$-coordinate being from $j$ to $j+1$ and $z$-coordinate being from $k$ to $k+1$, the left-hand side of Equation (4) is discretized as

$$
\frac{\phi^{\tau+1}_{i+\frac{1}{2},j+\frac{1}{2},k+\frac{1}{2}}C^{\tau+1}_{f,i+\frac{1}{2},j+\frac{1}{2},k+\frac{1}{2}}-\phi^{\tau}_{i+\frac{1}{2},j+\frac{1}{2},k+\frac{1}{2}}C^{\tau}_{f,i+\frac{1}{2},j+\frac{1}{2},k+\frac{1}{2}}}{\Delta t}
$$

$$
+\frac{u^{\tau+1}_{x,i+1,j+\frac{1}{2},k+\frac{1}{2}}C^{\tau+1}_{f,i+1,j+\frac{1}{2},k+\frac{1}{2}}-u^{\tau+1}_{x,i,j+\frac{1}{2},k+\frac{1}{2}}C^{\tau+1}_{f,i,j+\frac{1}{2},k+\frac{1}{2}}}{\Delta x}
$$

$$
+\frac{u^{\tau+1}_{y,i+\frac{1}{2},j+1,k+\frac{1}{2}}C^{\tau+1}_{f,i+\frac{1}{2},j+1,k+\frac{1}{2}}-u^{\tau+1}_{y,i+\frac{1}{2},j,k+\frac{1}{2}}C^{\tau+1}_{f,i+\frac{1}{2},j,k+\frac{1}{2}}}{\Delta y}
$$

$$
+\frac{u^{\tau+1}_{z,i+\frac{1}{2},j+\frac{1}{2},k+1}C^{\tau+1}_{f,i+\frac{1}{2},j+\frac{1}{2},k+1}-u^{\tau+1}_{z,i+\frac{1}{2},j+\frac{1}{2},k}C^{\tau+1}_{f,i+\frac{1}{2},j+\frac{1}{2},k}}{\Delta z},
$$

with the upwind scheme being used to get the concentration value on the face of the cube, since the computed concentration is imposed at the centre of the cube. If $\boldsymbol{D_e}$ is written as

$$\boldsymbol{D_e} = \begin{pmatrix} D_{xx} & D_{xy} & D_{xz} \\ D_{yx} & D_{yy} & D_{yz} \\ D_{zx} & D_{zy} & D_{zz} \end{pmatrix},$$

the first term of the right-hand side of Equation (4) can be discretized as

$$\frac{\frac{C_{f,i+\frac{3}{2},j+\frac{1}{2},k+\frac{1}{2}}^{\tau+1} - C_{f,i+\frac{1}{2},j+\frac{1}{2},k+\frac{1}{2}}^{\tau+1}}{\Delta x} * \phi_{i+1,j+\frac{1}{2},k+\frac{1}{2}}^{\tau+1} * D_{xx,i+1,j+\frac{1}{2},k+\frac{1}{2}}^{\tau+1} - \frac{C_{f,i+\frac{1}{2},j+\frac{1}{2},k+\frac{1}{2}}^{\tau+1} - C_{f,i-\frac{1}{2},j+\frac{1}{2},k+\frac{1}{2}}^{\tau+1}}{\Delta x} * \phi_{i,j+\frac{1}{2},k+\frac{1}{2}}^{\tau+1} * D_{xx,i,j+\frac{1}{2},k+\frac{1}{2}}^{\tau+1}}{\Delta x}$$

$$+ \frac{\overline{\frac{dC_f}{dy}}_{i+1,j+\frac{1}{2},k+\frac{1}{2}}^{\tau+1} * \phi_{i+1,j+\frac{1}{2},k+\frac{1}{2}}^{\tau+1} * D_{xy,i+1,j+\frac{1}{2},k+\frac{1}{2}}^{\tau+1} - \overline{\frac{dC_f}{dy}}_{i,j+\frac{1}{2},k+\frac{1}{2}}^{\tau+1} * \phi_{i,j+\frac{1}{2},k+\frac{1}{2}}^{\tau+1} * D_{xy,i,j+\frac{1}{2},k+\frac{1}{2}}^{\tau+1}}{\Delta x}$$

$$+ \frac{\overline{\frac{dC_f}{dz}}_{i+1,j+\frac{1}{2},k+\frac{1}{2}}^{\tau+1} * \phi_{i+1,j+\frac{1}{2},k+\frac{1}{2}}^{\tau+1} * D_{xz,i+1,j+\frac{1}{2},k+\frac{1}{2}}^{\tau+1} - \overline{\frac{dC_f}{dz}}_{i,j+\frac{1}{2},k+\frac{1}{2}}^{\tau+1} * \phi_{i,j+\frac{1}{2},k+\frac{1}{2}}^{\tau+1} * D_{xz,i,j+\frac{1}{2},k+\frac{1}{2}}^{\tau+1}}{\Delta x}$$

$$+ \frac{\overline{\frac{dC_f}{dx}}_{i+\frac{1}{2},j+1,k+\frac{1}{2}}^{\tau+1} * \phi_{i+\frac{1}{2},j+1,k+\frac{1}{2}}^{\tau+1} * D_{yx,i+\frac{1}{2},j+1,k+\frac{1}{2}}^{\tau+1} - \overline{\frac{dC_f}{dx}}_{i+\frac{1}{2},j,k+\frac{1}{2}}^{\tau+1} * \phi_{i+\frac{1}{2},j,k+\frac{1}{2}}^{\tau+1} * D_{yx,i+\frac{1}{2},j,k+\frac{1}{2}}^{\tau+1}}{\Delta y} +$$

$$\frac{\frac{C_{f,i+\frac{1}{2},j+\frac{3}{2},k+\frac{1}{2}}^{\tau+1} - C_{f,i+\frac{1}{2},j+\frac{1}{2},k+\frac{1}{2}}^{\tau+1}}{\Delta y} * \phi_{i+\frac{1}{2},j+1,k+\frac{1}{2}}^{\tau+1} * D_{yy,i+\frac{1}{2},j+1,k+\frac{1}{2}}^{\tau+1} - \frac{C_{f,i+\frac{1}{2},j+\frac{1}{2},k+\frac{1}{2}}^{\tau+1} - C_{f,i+\frac{1}{2},j-\frac{1}{2},k+\frac{1}{2}}^{\tau+1}}{\Delta y} * \phi_{i+\frac{1}{2},j,k+\frac{1}{2}}^{\tau+1} * D_{yy,i+\frac{1}{2},j,k+\frac{1}{2}}^{\tau+1}}{\Delta y}$$

$$+ \frac{\overline{\frac{dC_f}{dz}}_{i+\frac{1}{2},j+1,k+\frac{1}{2}}^{\tau+1} * \phi_{i+\frac{1}{2},j+1,k+\frac{1}{2}}^{\tau+1} * D_{yz,i+\frac{1}{2},j+1,k+\frac{1}{2}}^{\tau+1} - \overline{\frac{dC_f}{dz}}_{i+\frac{1}{2},j,k+\frac{1}{2}}^{\tau+1} * \phi_{i+\frac{1}{2},j,k+\frac{1}{2}}^{\tau+1} * D_{yz,i+\frac{1}{2},j,k+\frac{1}{2}}^{\tau+1}}{\Delta y}$$

$$+ \frac{\overline{\frac{dC_f}{dx}}_{i+\frac{1}{2},j+\frac{1}{2},k+1}^{\tau+1} * \phi_{i+\frac{1}{2},j+\frac{1}{2},k+1}^{\tau+1} * D_{zx,i+\frac{1}{2},j+\frac{1}{2},k+1}^{\tau+1} - \overline{\frac{dC_f}{dx}}_{i+\frac{1}{2},j+\frac{1}{2},k}^{\tau+1} * \phi_{i+\frac{1}{2},j+\frac{1}{2},k}^{\tau+1} * D_{zx,i+\frac{1}{2},j+\frac{1}{2},k}^{\tau+1}}{\Delta z}$$

$$+ \frac{\overline{\frac{dC_f}{dy}}_{i+\frac{1}{2},j+\frac{1}{2},k+1}^{\tau+1} * \phi_{i+\frac{1}{2},j+\frac{1}{2},k+1}^{\tau+1} * D_{zy,i+\frac{1}{2},j+\frac{1}{2},k+1}^{\tau+1} - \overline{\frac{dC_f}{dy}}_{i+\frac{1}{2},j+\frac{1}{2},k}^{\tau+1} * \phi_{i+\frac{1}{2},j+\frac{1}{2},k}^{\tau+1} * D_{zy,i+\frac{1}{2},j+\frac{1}{2},k}^{\tau+1}}{\Delta z} +$$

$$\frac{C_{f,i+\frac{1}{2},j+\frac{1}{2},k+\frac{3}{2}}^{\tau+1} - C_{f,i+\frac{1}{2},j+\frac{1}{2},k+\frac{1}{2}}^{\tau+1}}{\Delta z} * \phi_{i+\frac{1}{2},j+\frac{1}{2},k+1}^{\tau+1} * D_{zz,i+\frac{1}{2},j+\frac{1}{2},k+1}^{\tau+1} - \frac{C_{f,i+\frac{1}{2},j+\frac{1}{2},k+\frac{1}{2}}^{\tau+1} - C_{f,i+\frac{1}{2},j+\frac{1}{2},k-\frac{1}{2}}^{\tau+1}}{\Delta z} * \phi_{i+\frac{1}{2},j+\frac{1}{2},k}^{\tau+1} * D_{zz,i+\frac{1}{2},j+\frac{1}{2},k}^{\tau+1}}{\Delta z}.$$

The notation $\overline{\frac{dC_f}{dy}}_{i+1,j+\frac{1}{2},k+\frac{1}{2}}^{\tau+1}$ stands for the average value of $\frac{dC_f}{dy}$ on the $x$-direction face with its $x$-coordinate being $i+1$, which is calculated by the four $\frac{dC_f}{dy}$ on the four $y$-direction faces adjacent to the $x$-direction face. The meanings of the other similar notations are analogous. The discretization of the second term of the right-hand side of Equation (4) is trivial, which is not given here. It is easy to see that the stencil pattern of $T$ is the same as $C_f$, and thus the discretization of Equation (21) holds the same philosophy as that of Equation (4), and the details are not given any more.

In order to use the experimenting field approach to compute the coefficients of the three linear systems, the unknowns to be computed can be divided into four fields: the velocity field, the pressure field, the concentration field, and the temperature field. If there is a 3D domain and it can be divided into eight cubes as shown in Figure 6, and then the velocity field can be represented as arrows on the faces of the cubes, and the pressure field, concentration field and temperature field can be represented as points at the centres of the cubes. Each $x$-momentum, $y$-momentum, and $z$-momentum conservation equation can be discretized on each $x$-direction, $y$-direction, and $z$-direction face, respectively. Each mass conservation equation, concentration balance equation, and energy conservation equation can be discretized at each centre of the cube. Such kind of grid is called a staggered grid in CFD. The experimenting field approach used in the 2D simulation [19] is expanded to the 3D case directly, and the details are not given in this work anymore.

In order to capture the details of the configuration of the matrix after acidization, a fine 3D grid is needed in the simulation, which brings about a huge number of cells in the 3D grid, and as a result of that parallelisation is necessary to be introduced in the simulation. At the first step, domain decomposition has to be done on the 3D domain. The main purpose of domain decomposition is to allocate the discretized equations to the processors. Suppose there is a 3D Cartesian grid with $nx$, $ny$, and $nz$ cubes in the $x$-, $y$-, and $z$-direction, respectively, and there are $npx$, $npy$, and $npz$ processors in the $x$-, $y$-, and $z$-direction, respectively. $nx$, $ny$, and $nz$ are supposed to be divisible by $npx$, $npy$, and $npz$, respectively. Furthermore, it is stipulated that $\frac{nx}{npx} \geq 2$, $\frac{ny}{npy} \geq 2$, and $\frac{nz}{npz} \geq 2$. For the processor with the coordinate ($I,J,K$), the following equations discretized at the centres of the cubes with the coordinate ($i,j,k$) are allocated to it

$$(I-1) * \frac{nx}{npx} + 1 \leq i \leq I * \frac{nx}{npx},$$

$$(J-1) * \frac{ny}{npy} + 1 \leq j \leq J * \frac{ny}{npy},$$

$$(K-1) * \frac{nz}{npz} + 1 \leq k \leq K * \frac{nz}{npz},$$

$$1 \leq I \leq npx, 1 \leq J \leq npy, 1 \leq K \leq npz,$$

and the equations discretized on the *x*-direction faces with the coordinate $(i,j,k)$, which are the *x*-momentum conservation equations, are allocated to it

$$(I-1)*\frac{nx}{npx}+1 \leq i \leq I*\frac{nx}{npx}+\delta_x,$$

$$(J-1)*\frac{ny}{npy}+1 \leq j \leq J*\frac{ny}{npy},$$

$$(K-1)*\frac{nz}{npz}+1 \leq k \leq K*\frac{nz}{npz},$$

$$\delta_x = \begin{cases} 1, & if\ I=npx \\ 0, & otherwise \end{cases},$$

$$1 \leq I \leq npx, 1 \leq J \leq npy, 1 \leq K \leq npz.$$

The allocation strategy of the *y*-momentum conservation equations and the *z*-momentum conservation equations is similar. After the allocation of the discretized equations, the variables that are needed by the equations should also be allocated to the processors, with some variables being communicated among the processors. By the domain decomposition strategy, each discretized equation can be allocated to only one processor, with the benefits that it can keep load balance of the processors and reduce the communication cost among the processors.

After domain decomposition, a suitable parallel solver can be leveraged to solve the three linear systems. In the 2D parallel simulation of the work [19], the parallel solver HYPRE is used, but it can only solve few simple cases. As a result of that, more complicated cases have to be solved by the direct solver UMFPACK [40] in a serial code, which limits the application of the parallel code. In this work, another parallel solver called MUMPS is used, which can solve the complicated cases that HYPRE cannot solve. With the help of MUMPS, the improved DBF framework and the thermal DBF framework are parallelized successfully for the first time, which makes the fine 3D simulation feasible. It is emphasized that different from HYPRE that is an iterative solver, MUMPS is a direct solver, which makes the time step can be set larger in the simulation. Moreover, the direct solver can solve the linear system directly, with no need to add a pseudo parameter $\varepsilon$ in the mass conservation equation, which is the case in the work [19]. There, $\varepsilon$ is introduced to ensure a linear system with an invertible coefficient matrix; otherwise the iterative solver HYPRE cannot solve it. However, the introduction of $\varepsilon$ changes the attribute of the flow in matrix acidization from incompressible to a little compressible, which contradicts the real case and makes the DBF framework less reliable. Lastly, with the help of FORTRAN90 and MPI, a series of 2D and 3D parallel codes are developed. In the following sections, these codes are used to run a series of numerical experiments on the supercomputer Shaheen [41].

## 5. Verification of the Improved DBF Framework

**5.1 3D shear-driven cavity flows**

It is learnt that the model of shear-driven cavity flows [42] is in fact a reduction of the improved DBF framework, since Equation (2) and (3) reduce to the following two equations, respectively,

$$\frac{\partial \boldsymbol{u}}{\partial t} + \nabla \cdot \boldsymbol{u}\boldsymbol{u} = -\nabla p + \frac{1}{Re}\nabla^2\boldsymbol{u},$$

$$\nabla \cdot \boldsymbol{u} = 0,$$

which describes shear-driven cavity flows. Therefore, the 3D code of the improved DBF framework can be used to simulate 3D shear-driven cavity flows all the same, as long as some parameters are simplified, such as ignoring the concentration balance equation. In this issue of 3D shear-driven cavity flows, a laminar incompressible flow is inside a unit cube cavity whose $y$-direction top surface is moved by an $x$-direction uniform velocity of 1 m/s, as shown in Figure 7. The Reynolds number ($Re$) is 100. The gravity effect is ignored. The grid has $20^3$ cubes. The simulation results of stable flows are shown in Figure 8 and Figure 9, respectively. The two figures display the velocity profiles of the $x$-direction component on the vertical centreline and the $y$-direction component on the horizontal centreline of the plane $z = 0.5$, respectively. The simulation results can be compared with Fig. 6 in [42]. To the eyeball resolution, we cannot clarify their differences, which proves the correctness of the 3D code of the improved DBF framework to some extent.

**5.2 2D linear flows**

The 2D linear flows in the work [17] are simulated again by the improved DBF framework, with more or less the same experimental parameters, which are shown in Table 2. In [17], the flows are simulated with the Darcy framework [15], but Navier-Stokes fluid dynamics is considered. Thus, among all the state-of-the-art models developed from the Darcy framework, the model is close to the improved DBF framework, and its results can be compared with the results from the improved DBF framework. It is noted that the subscript 0 represents the initial value. $\bar{\phi}_0$ represents the initial average porosity in the medium, with heterogeneity magnitude of 0.03. The gravity effect is ignored. In the 2D simulation, there is a rectangular matrix of 0.1-meter length ($x$-direction) and 0.04-meter width ($y$-direction). Acid flow is injected into the matrix from the left boundary and goes out of the matrix from the right boundary, which means that the injected velocity is imposed on the left boundary, and Dirichlet boundary condition for pressure is imposed on the right boundary. It is stipulated that the pressure imposed on the right boundary is the same as the initial pressure in the matrix. For concentration, Dirichlet boundary condition is imposed on the left boundary, and no-flux boundary condition is imposed on the right boundary. The upper and lower boundaries are closed for both pressure and concentration, which means no-flow, no-flux boundary conditions are imposed. Acid concentration is zero in the matrix initially. The injected velocity of the acid flow of 0.5 M hydrogen chloride (HCl) on the left boundary is changed in the simulation, which leads to different configurations of the matrix after acidization. The grid has 180 cells in the $x$-direction and 72 cells in the $y$-direction, which is the same size as that of [17]. Since the work [17] declares their grid is fine enough to describe matrix acidization, our grid is also capable to do that.

The pore volumes to breakthrough (PVBT) of different injected velocities are given in Figure 10, and the curve in the figure is called the acid-efficiency curve. Breakthrough is defined as the moment when the pressure drop across the medium drops to 1% of its initial value [43]. From Figure 10, it can be seen that the acid-efficiency curve in this work matches the corresponding curve in Fig. 8 of [17] well when the range of the injected velocity is from $4.17 \times 10^{-7}$ m/s to $1.67 \times 10^{-4}$ m/s. When the injected velocity is $1.67 \times 10^{-7}$ m/s, its corresponding PVBT is 6.62, which is not reasonable. The injected velocity of $1.67 \times 10^{-7}$ m/s is very slow, which may indicate the face dissolution pattern. According to the work [16], in such condition, about 100 million grid cells are needed to capture the face dissolution pattern accurately, which indicates that our grid is not fine enough to output accurate results. Unfortunately, due to the limit of the supercomputing power, the simulation based on the 100 million grid cells cannot be finished in a reasonable time, and so it is not done in this work. The minimum PVBT is 4.54, which is achieved at the injected velocity of $4.17 \times 10^{-6}$ m/s.

The fixed time step is assumed for the simulations, and the time steps corresponding to the injected velocities are shown in Table 3. It is emphasized that all the time steps make sure the Courant number be less than one. The sensitivity test is done when each of the time steps is increased by two times and still makes sure the Courant number be less than one, and all the values of PVBT are the same, which demonstrates the simulation results in Figure 10 can be deemed as the true results.

The porosity profiles at breakthrough corresponding five different injected velocities are given in Figure 11. From the figure, it can be seen that five dissolution patterns appear in their turns when the injected velocity increases.

### 5.3 3D linear flows

The simulation of 2D linear flows has achieved reasonable results, and it can be expanded to the simulation of 3D linear flows by given another dimension to the matrix, with the length of 0.04-meter. After this, a 3D matrix comes with 0.1-meter length in the $z$-direction, 0.04-meter length in the $x$- and $y$-direction, respectively. According to the work [17], dissolution patterns from the conical wormhole to the uniform dissolution can be captured accurately only when the grid has at least 180 cells in the $z$-direction and 72 cells in the $x$-direction and $y$-direction, respectively. Meanwhile, the face dissolution pattern requires a finer grid than that tone. However, to make sure the Courant number be less than one, the number of iteration steps can be huge, which brings about a long simulation period. Even though the code runs on Shaheen, at least one month is needed for the fastest case to breakthrough. Thus, that kind of grid is beyond the computing ability currently, and a coarser grid is given in this work. The number of cells is divided by two in each dimension, and a coarser grid, with 90 cells in the $z$-direction and 36 cells in the $x$-direction and $y$-direction, respectively, is used to simulate the 3D linear flows. Although the coarser grid is not fine enough to capture all kinds of dissolution patterns accurately, its simulation results can still be used to verify the correctness of the 3D code of the improved DBF framework. The acid flow is injected into the matrix along the $z$-direction. The other experimental parameters and boundary conditions are the same as the 2D simulation.

The time steps for different injected velocities are shown in Table 4. There are two groups of time steps, with the values of the second column being two times of the corresponding values of the third column, the purpose of which is to test the convergence of the results. All the time steps guarantee the Courant number is less than one. It is emphasized that due to the limit of Shaheen, a code can run on it for at most three days. Therefore, the simulations for the velocities $3.04 \times 10^{-7}$ m/s and $1.04 \times 10^{-4}$ m/s at fine time steps are not done, since their simulation time to achieve breakthrough is beyond three days. The acid-efficiency curves for both groups of time steps are shown in Figure 12. The numbers beside the points represent the values of PVBT, with the blue ones coming from coarse time steps and the red ones coming from fine time steps. From the figure, it can be learnt that the values of PVBT from coarse time steps are very near from those from fine time steps, which demonstrates the convergence of the results. Moreover, the values of PVBT tend to decrease with finer time steps. It can be expected that with finer grids, smaller PVBT values can be achieved. The minimum PVBT is 3.567, which is achieved at the injected velocity of $1.04 \times 10^{-6}$ m/s. This coincides with the work [17] where the minimum PVBT is achieved at the injected velocity of 0.1 cm$^3$/min which is the same as $1.04 \times 10^{-6}$ m/s. However, in [17], the minimum PVBT is about two that is smaller than ours, which is due to finer grid. Except the points at the injected velocity of $1.04 \times 10^{-7}$ m/s, the shape of the acid-efficiency curves matches the corresponding one in [17] well. The drop of the PVBT values at that velocity is due to the inaccuracy of the simulation when the grid is not fine enough, which can also be seen in the simulation of 2D linear flows above. Furthermore, the minimum PVBT of 2D simulations is larger than the one of 3D simulations, and the injected velocity of 2D simulations at which the minimum PVBT is achieved is also larger than that of 3D simulations, which conforms to the qualitative trends in [15].

These effects of the injected velocities on dissolution patterns are shown in Figure 13. From the figure, it can be seen that five different dissolution patterns can be simulated.

## 6. Verification of the Thermal DBF Framework

### 6.1 Isothermal conditions

The correctness of the improved DBF framework is a major premise of the thermal DBF framework, which has been verified in last section. The correctness of the thermal DBF framework is discussed in this section. Firstly, an experiment is carried out, with isothermal conditions in which the injected acid temperature and the initial matrix temperature are the same. Since 2D experiments are eligible to verify the correctness of the model, the grid of the 2D linear flows is used again, with 180×72 cells totally. 3D experiments are left to future work. In order to compare the numerical results with the chemical results of the "effects of temperature" experiment of the work [31], three kinds of temperature are chosen: 295K, 323K, and 353 K, which corresponds to 22 ℃, 50 ℃, and 80 ℃ in [31]. The boundary conditions and initial conditions for pressure and concentration are the same as those of the 2D linear flow experiment above. Besides that, for temperature, adiabatic conditions are applied, which means that except the acid injection boundary (left boundary), all the other boundaries are adiabatic ones. The experimental parameters can be seen in Table 2. It is noted that the values of $d_m$ and $k_s$ in Table 2 are not used in the experiments of this section, since they are variables in the thermal DBF framework. All the parameters are more or less the same as those in the "effects of temperature" experiment of [31].

The values of PVBT for different injected velocities and temperatures are shown in Table 5. From the table, it can be known that when the temperature is 295 K, the minimal PVBT is 4.350 which is achieved at the optimal injected velocity of $2.67 \times 10^{-6}$ m/s; when the temperature is 323 K, the minimal PVBT is 4.362 which is achieved at the optimal injected velocity of $9.17 \times 10^{-6}$ m/s; when the temperature is 353 K, the minimal PVBT is 4.416 which is achieved at the optimal injected velocity of $4.17 \times 10^{-5}$ m/s. It is emphasized that in order to verify the convergence of the results, both coarse-time-step and fine-time-step results are computed. The coarse-time step is two times of the fine-time step for every injected velocity. All the time steps can guarantee the Courant number is less than one. From the table, it can be learnt that the differences of the coarse-time-step and fine-time-step results are very little, which means convergence is achieved. The values of PVBT from fine-time steps constitute the acid-efficiency curves of different temperatures in Figure 14. Since these experimental parameters hint that the injected acid is 0.5 M HCl and the matrix is limestone, Figure 14 is compared with Fig. 6 of [31]. From the two figures, it is evident that both of the minimal PVBT and the optimal injected velocity increase when the temperature increases, which means that the numerical simulation results can be observed in labs. It is noted that when the injected velocity is below about $4.17 \times 10^{-6}$ m/s, the values of PVBT increase with the increase of temperature; when the injected velocity is above about $4.17 \times 10^{-5}$ m/s, the values of PVBT decrease with the increase of temperature, which means the former one is a mass-transfer controlled regime and the latter one is a kinetically controlled regime. Furthermore, for the three different temperatures, the porosity profiles at breakthrough in the optimal injected velocity are given in Figure 15. From the figure, it can be seen that with the increase of the temperature, the diameter of the wormhole also increases, which matches the observation from Fig. 7 of [31]. This also explains why the minimal PVBT value increases with the increase of temperature. In fact, the transferring efficiency of HCl decreases due to increased acid consumption on the walls of the wormhole, which brings about the phenomenon above.

## 6.2 Non-isothermal conditions

Isothermal conditions are common in labs. However, in field cases, the injected acid temperature is often different from the initial matrix temperature. Due to the geothermal factor, the initial matrix temperature may be higher than the injected acid temperature. Thus, in order to expand the simulations from labs to fields, non-isothermal conditions are considered. In the experiment, four cases with different combinations of the injected acid temperature and initial matrix temperature are simulated, with the injected velocity of $9.17 \times 10^{-6}$ m/s, and the values of PVBT are shown in Table 6. The other experimental parameters are the same as the isothermal conditions. All the results are from the fine-time-step simulations. From Table 6, it can be seen that when the injected acid temperature is fixed at 323 K, the changed initial matrix temperatures will not change the values of PVBT. For three different initial matrix temperatures: 295 K, 323 K, and 353 K, the values of PVBT are the same as 4.362. However, when the injected acid temperature is changed, even though the initial matrix temperature is unchanged, the values of PVBT will be changed, which is evident from the first two rows of Table 6. From the two rows, it can be further found that the two PVBT values are nearly the same as the corresponding values in the isothermal conditions, respectively, which also demonstrates the injected acid temperature, instead of the initial matrix temperature, has effect on the PVBT value. Thus, the injected acid temperature governs the PVBT value, and can be a design parameter in matrix acidization, which can also be concluded from [30]. The porosity profiles at breakthrough for three different initial matrix temperatures are given in Figure 16, respectively, where the same kind of porosity profiles can be seen clearly. Thus, Figure 16 demonstrates the temperature of the initial matrix is not a key factor to affect matrix acidization once again.

In order to learn the reason why the injected acid temperature has such a significant effect on matrix acidization, the change of the average matrix temperature with time is investigated. The initial matrix temperature is set as 323 K, and the injected acid temperatures are 295 K and 353 K, respectively, which are also the cases represented by the first two rows of Figure 6. The history of the average matrix temperature from the beginning to breakthrough is shown in Figure 17. From the figure, it is learnt that the matrix temperature becomes more or less the same as the injected acid temperature immediately after acidization begins, and continues to be like this till breakthrough. This explains why the initial matrix temperature almost has no effect on matrix acidization.

After the discussions above, it is concluded that the thermal DBF framework can simulate reasonable numerical results which are verified by the numerical and chemical experiments of the other works. This indicates that the thermal DBF framework can be an effective tool in the field of matrix acidization.

# 7. Performance Evaluation

The performance of the 2D parallel code has been evaluated in [19], and this work tries to evaluate the performance of the 3D parallel code. The test is done on the 3D grid used above. Moreover, the number of the iterations is set smaller to 100 to save the supercomputing resources. Meanwhile, since the sparsity pattern of the coefficient matrix of the linear system of the energy conservation equation is the same as that of the concentration balance equation, evaluating both of the linear systems is a redundancy. Therefore, only the linear system of the concentration balance equation is evaluated, which brings about the performance of the 3D parallel code of the improved DBF framework is evaluated in this section. Its evaluation result is trustable to foresee the performance of the 3D parallel code of the thermal DBF framework.

The experiment of the improved DBF framework in the verification section is done again, and the performance results are shown in Figure 18. From the figure, it can be seen that the solver time takes up most part of the run time, which is reasonable. Moreover, the run time decreases with the increase of the number of processors, which means some kind of speedup can be achieved. However, when the number of processors increases to 144, further speedup seems impossible. The similar scalability of the solver MUMPS can be seen in the work [44].

# 8. Conclusion and Future Work

Since Wu et al. [19] contributed the DBF framework to the field of matrix acidization, polish to the framework is always on the way. This work is one of the endeavours, and tries to correct a defect in the momentum conservation equation of the DBF framework and make the momentum conservation equation still reliable when porosity is changed. Furthermore, by introducing a direct solver called MUMPS, the pseudo parameter $\varepsilon$ in the mass conservation equation can be deleted, which keeps the incompressible attribute of the acid flow in matrix acidization and thus makes the framework more reasonable. Besides that, the simulation flowchart is also changed in this work, which is another correction to the DBF framework. After these polishes, the new framework can be called the improved DBF framework for short. The improved DBF framework is realized by 2D and 3D parallel codes with the help of MPI and FORTRAN 90, and verified by comparison with a series of former works. It is emphasized that the 3D simulation results of the improved DBF framework are given for the first time in this work. The improved DBF framework can simulate similar numerical results with the works [42] and [17], which demonstrates its reliability.

The correctness of the improved DBF framework makes it feasible to develop the thermal DBF framework based on it. Besides the mass conservation law and momentum conservation law which are included in the improved DBF framework, the thermal DBF framework also considers the energy conservation law and thus introduces the energy balance equation to the improved DBF framework. Verification to the thermal DBF framework is done on isothermal conditions and non-isothermal conditions, respectively, and the numerical simulation results match the conclusions from the other chemical and numerical experiments such as [31] and [30] well. Therefore, the thermal DBF framework is reasonable and trustable.

Since the accuracy of matrix acidization simulation depends on the size of the grid deeply, very fine grids are required by trustable results, which brings about the need to develop parallel codes to finish simulations in reasonable time. However, for the improved DBF framework and thermal DBF framework, parallelizing them is not an easy task, due to their complex equation systems. With the help of MPI and FORTRAN 90, and the experimenting field approach, this work overcomes the difficulties and develops good-scalability parallel codes, which is another big contribution to the field of matrix acidization study.

With the reliable improved DBF framework and thermal DBF framework, a series of numerical investigations on matrix acidization can be carried out in the future, and more reasonable results are expected.

## References

[1]	Liu, Piyang, et al. "Modeling and simulation of wormhole formation during acidization of fractured carbonate rocks." Journal of Petroleum Science and Engineering 154 (2017): 284-301.

[2]	Ma, Guowei, et al. "Modelling temperature-influenced acidizing process in fractured carbonate rocks." International Journal of Rock Mechanics and Mining Sciences 105 (2018): 73-84.

[3]	Szymczak, Piotr, et al. "Wormhole Formation During Acidizing of Calcite-Cemented Fractures in Gas-Bearing Shales." SPE Journal (2019).

[4]	Zhao C, Hobbs BE, Ord A, Hornby P, Peng S. Effect of reactive surface areas associated with different particle shapes on chemical-dissolution front instability in fluid-saturated porous rocks. Transp Porous Media 2008;73:75-94.

[5]	Zhao C, Hobbs BE, Ord A, Peng S. Effects of mineral dissolution ratios on chemical-dissolution front instability in fluid-saturated porous media. Transp Porous Media 2010;82:317-35.

[6]	Zhao C, Hobbs BE, Ord A. Theoretical analyses of the effects of solute dispersion on chemical-dissolution front instability in fluid-saturated porous media. Transp Porous Media 2010;84:629-53.

[7]	Buijse MA. Understanding wormholing mechanisms can improve acid treatments in carbonate formations. SPE Prod Facilities 2000;15:168-75.

[8]	Hoefnger ML, Fogler HS. Pore evolution and channel formation during flow and reaction in porous media. AIChE J 1988;34:45-54.

[9]	Fredd CN, Fogler HS. Influence of transport and reaction on wormhole formation in carbonate porous media. AIChE J 1998;44:1933-49.

[10]	Daccord G, Touboul E, Lenormand R.. Carbonate acidizing: toward a quantitative model of the wormholing phenomenon. SPE Prod Eng 1989;63-8.

[11]	Daccord, G., R. Lenormand, and O. Lietard. "Chemical dissolution of a porous medium by a reactive fluid—I. Model for the "wormholing" phenomenon."Chemical Engineering Science 48.1 (1993): 169-178.

[12]	Daccord, G., O. Lietard, and R. Lenormand. "Chemical dissolution of a porous medium by a reactive fluid—II. Convection vs reaction, behavior diagram."Chemical engineering science 48.1 (1993): 179-186.

[13]	Liu X, Ormond A, Bartko K, Ortoleva P. A geochemical reaction-transport simulator for matrix acidizing analysis and design. J Pet Sci Eng 1997;17:181-97.

[14]	Golfier F, Zarcone C, Bazin B, Lenormand R, Lasseux D, Quintard M. On the ability of a Darcy-scale model to capture wormhole formation during the dissolution of a porous medium. J Fluid Mech 2002;457:213-54.

[15]   Panga MKR, Ziauddin M, Balakotaiah V. Two-scale continuum model for simulation of wormholes in carbonate acidization. AIChE J 2005;51:3231-48.

[16]   Maheshwari, Priyank, and Vemuri Balakotaiah. "Comparison of carbonate HCl acidizing experiments with 3D simulations." SPE Production & Operations 28.04 (2013): 402-413.

[17]   Akanni, Olatokunbo O., Hisham A. Nasr-El-Din, and Deepak Gusain. "A Computational Navier-Stokes Fluid-Dynamics-Simulation Study of Wormhole Propagation in Carbonate-Matrix Acidizing and Analysis of Factors Influencing the Dissolution Process." SPE Journal 22.06 (2017): 2-049.

[18]   De Oliveira, Thiago Judson Lima, et al. "Numerical simulation of the acidizing process and PVBT extraction methodology including porosity/permeability and mineralogy heterogeneity." SPE International Symposium and Exhibition on Formation Damage Control. Society of Petroleum Engineers, 2012.

[19]   Wu, Yuanqing, Amgad Salama, and Shuyu Sun. "Parallel simulation of wormhole propagation with the Darcy–Brinkman–Forchheimer framework." Computers and Geotechnics 69 (2015): 564-577.

[20]   Wu, Yuanqing. Parallel Reservoir Simulations with Sparse Grid Techniques and Applications to Wormhole Propagation. Diss. 2015.

[21]   Kou, Jisheng, Shuyu Sun, and Yuanqing Wu. "Mixed finite element-based fully conservative methods for simulating wormhole propagation." Computer Methods in Applied Mechanics and Engineering 298 (2016): 279-302.

[22]   Kou, Jisheng, Shuyu Sun, and Yuanqing Wu. "A semi-analytic porosity evolution scheme for simulating wormhole propagation with the Darcy–Brinkman–Forchheimer model." Journal of Computational and Applied Mathematics 348 (2019): 401-420.

[23]   Li, Xiaoli, and Hongxing Rui. "Superconvergence of a fully conservative finite difference method on non-uniform staggered grids for simulating wormhole propagation with the Darcy–Brinkman–Forchheimer framework." Journal of Fluid Mechanics 872 (2019): 438-471.

[24]   Li, Xiaoli, Hongxing Rui, and Shuangshuang Chen. "A fully conservative block-centered finite difference method for simulating Darcy-Forchheimer compressible wormhole propagation." Numerical Algorithms 82.2 (2019): 451-478.

[25]   Li, Xiaoli, and Hongxing Rui. "A fully conservative block‐centered finite difference method for Darcy‐Forchheimer incompressible miscible displacement problem." Numerical Methods for Partial Differential Equations 36.1 (2020): 66-85.

[26]   Falgout, Robert D., and Ulrike Meier Yang. "hypre: A library of high performance preconditioners." International Conference on Computational Science. Springer, Berlin, Heidelberg, 2002.

[27]   Amestoy, Patrick R., et al. "A fully asynchronous multifrontal solver using distributed dynamic scheduling." SIAM Journal on Matrix Analysis and Applications 23.1 (2001): 15-41.

[28]   Amestoy, Patrick R., et al. "Hybrid scheduling for the parallel solution of linear systems." Parallel computing 32.2 (2006): 136-156.

[29]   Li, Yongming, et al. "Simulation and analysis of wormhole formation in carbonate rocks considering heat transmission process." Journal of Natural Gas Science and Engineering 42 (2017): 120-132.

[30]   Kalia, Nitika, and Gerard Glasbergen. "Fluid temperature as a design parameter in carbonate matrix acidizing." SPE Production and Operations Conference and Exhibition. Society of Petroleum Engineers, 2010.

[31]    Fredd, C. N., and H. S. Fogler. "Optimum conditions for wormhole formation in carbonate porous media: Influence of transport and reaction." Spe Journal 4.03 (1999): 196-205.

[32]    Whitaker, Stephen. "The Forchheimer equation: a theoretical development." Transport in Porous media 25.1 (1996): 27-61.

[33]    Aavatsmark, Ivar. "An introduction to multipoint flux approximations for quadrilateral grids." Computational Geosciences 6.3 (2002): 405-432.

[34]    Sekutkovski, Bojan, et al. "A hybrid RANS-LES method with compressible k-omegaSSTSAS turbulence model for high Reynolds number flow applications/Hibridna RANS-LES metoda s kompresibilnim k-omegaSSTSAS turbulentnim modelom namjenjena analizi strujanja pri velikim Reynoldsovim brojevima." Tehnicki Vjesnik-Technical Gazette 22.5 (2015): 1237-1246.

[35]    Sun S, Salama A, El-Amin MF. An equation-type approach for the numerical solution of the partial differential equations governing transport phenomena in porous media. In: The International Conference on Computational Science, ICCS, June 4-6, Omaha, NE, 2012.

[36]    Salama A, Sun S, El Amin MF. A multi-point flux approximation of the steady state heat conduction equation in anisotropic media. ASME J Heat Transfer 2013;135:1-6.

[37]    Salama A, Li W, Sun S. Finite volume approximation of the three- dimensional flow equation in axisymmetric, heterogeneous porous media based on local analytical solution. J Hydrol 2013;501:45-55.

[38]    Salama A, Sun S, Wheeler M. Solving global problem by considering multitude of local problems: application to flow in anisotropic porous media using the multipoint flux approximation. J Comput Appl Math 2014;267:117-30.

[39]    Salama A, Sun S, El Amin MF. Investigation of thermal energy transport from an anisotropic central heating element to the adjacent channels: a multipoint flux approximation. Ann Nucl Energy 2015;76:100-12.

[40]    Davis TA. Algorithm 832: UMFPACK V4. 3---an unsymmetric-pattern multifrontal method. ACM Trans Math Software 2004;30:196-9.

[41]    Hadri, Bilel, et al. "Overview of the KAUST's Cray X40 system–Shaheen II." Proceedings of the 2015 Cray User Group (2015).

[42]    Ku, Hwar C., Richard S. Hirsh, and Thomas D. Taylor. "A pseudospectral method for solution of the three-dimensional incompressible Navier-Stokes equations." Journal of Computational Physics 70.2 (1987): 439-462.

[43]    Kalia, Nitika, and Vemuri Balakotaiah. "Effect of medium heterogeneities on reactive dissolution of carbonates." Chemical Engineering Science 64.2 (2009): 376-390.

[44]    Raju, Mandhapati P. "Parallel computation of finite element Navier-Stokes codes using MUMPS solver." arXiv preprint arXiv:0910.1845 (2009).

**Figures**

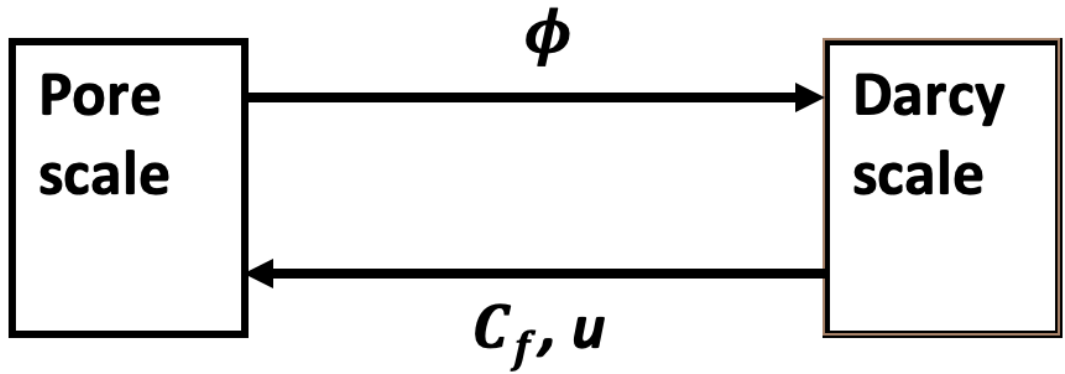

Figure 1 The interaction of the pore scale and Darcy scale. The upper arrow means that the pore scale will affect Darcy scale by the pore-scale variable $\phi$. As long as $\phi$ is changed, the pore-scale variables used in the Darcy scale are changed, which brings about the changes of the main variables $\boldsymbol{u}$, $p$, and $C_f$ in the Darcy scale. The lower arrow means that the Darcy scale will affect the pore scale by the Darcy-scale variables $C_f$ and $\boldsymbol{u}$, since the two variables will affect the value of $\phi$ directly. Once $\phi$ is changed, all the other variables in the pore scale will be changed accordingly.

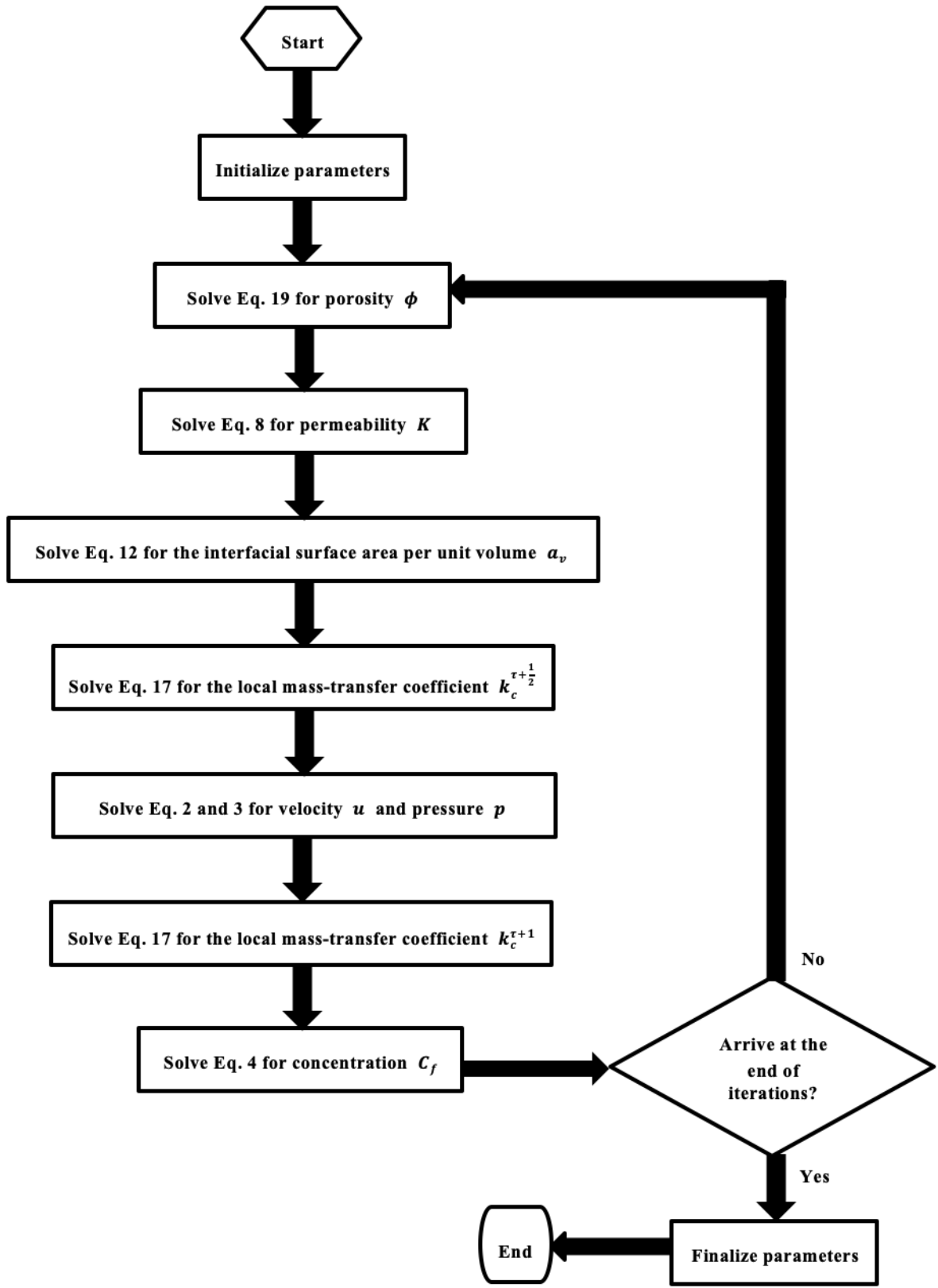

Figure 2 The flowchart of the improved DBF framework.

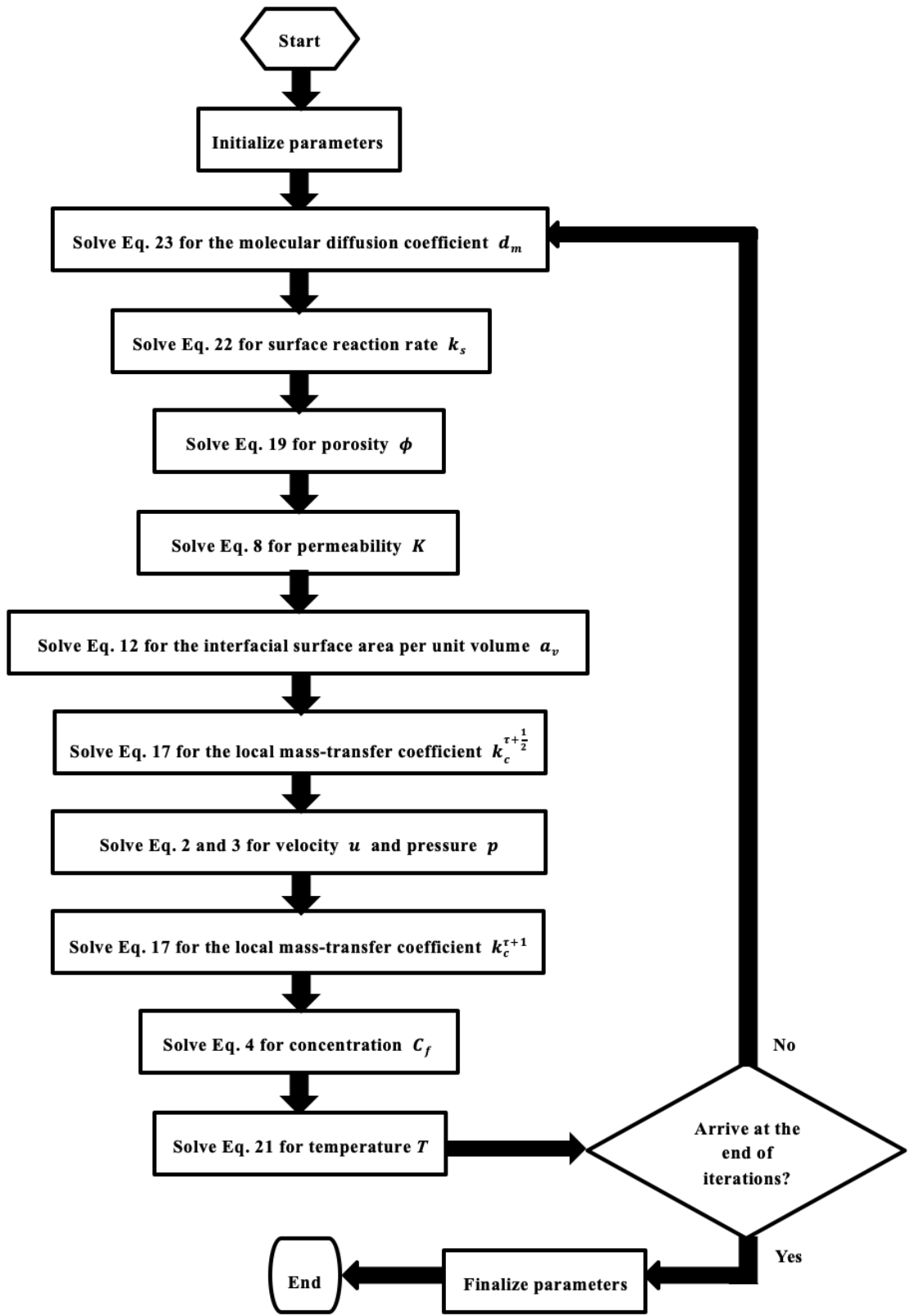

Figure 3 The flowchart of the thermal DBF framework.

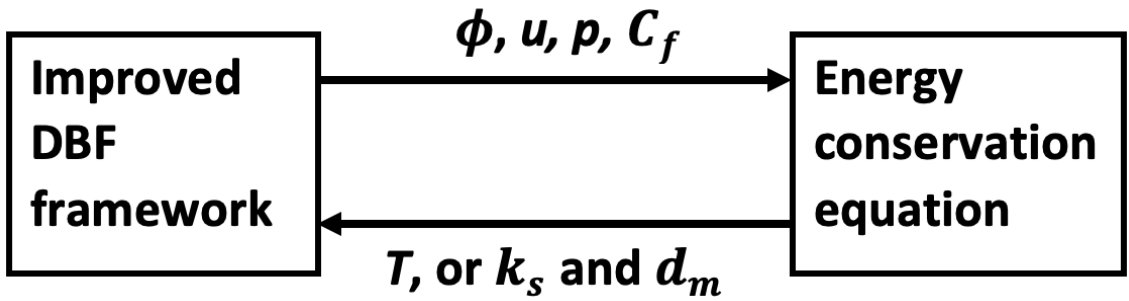

Figure 4 The interaction of the improved DBF framework and the energy conservation equation.

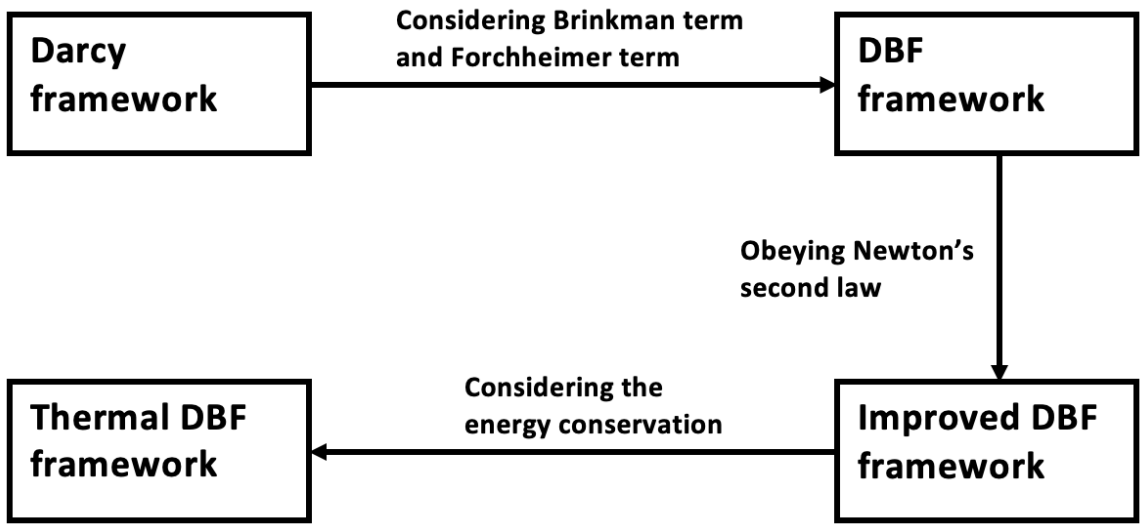

Figure 5 The relationship among the Darcy framework, the DBF framework, the improved DBF framework, and the thermal DBF framework.

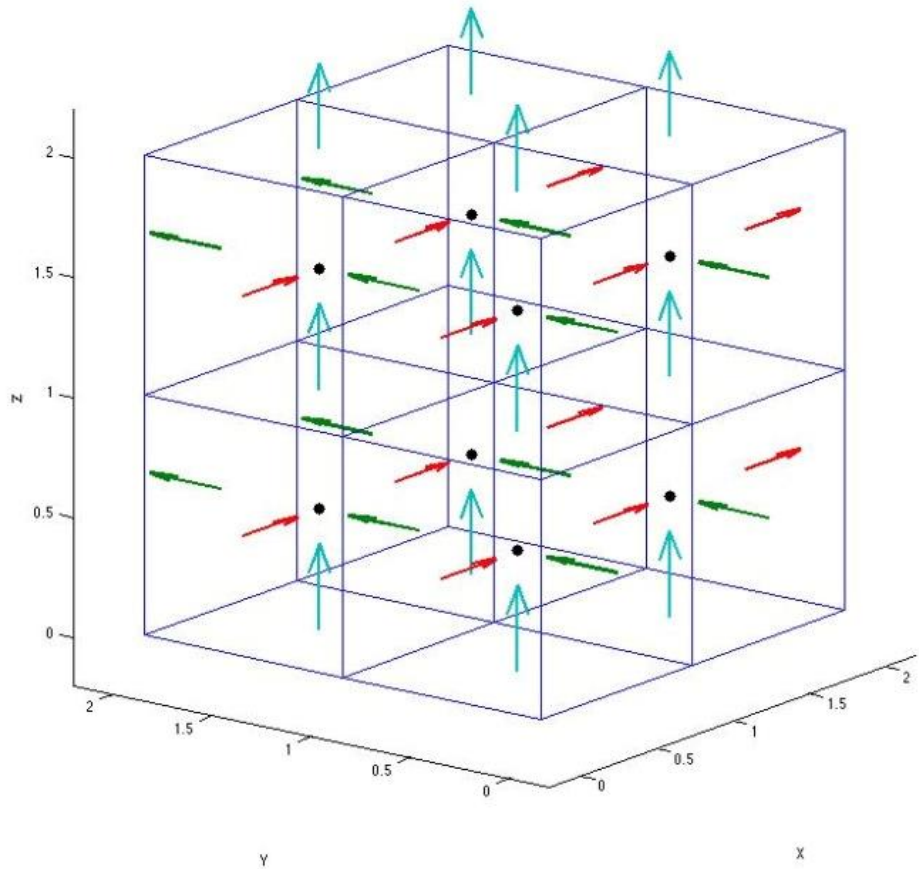

Figure 6 Staggered grid.

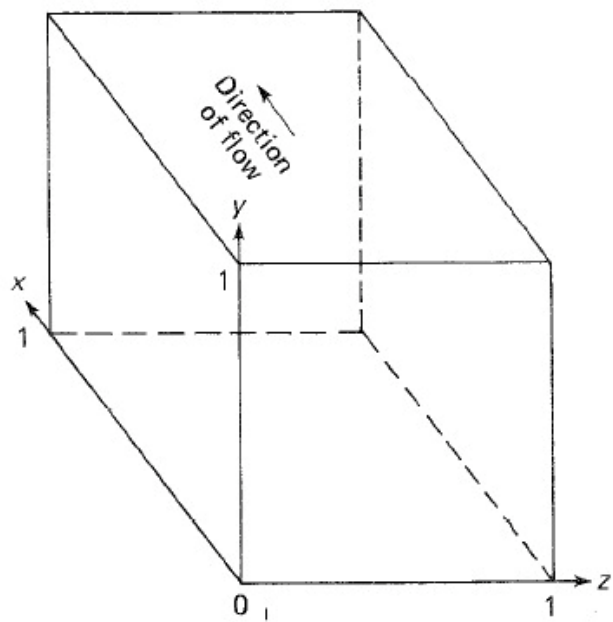

Figure 7 3D Shear-driven cavity flow configuration and coordinate system [42].

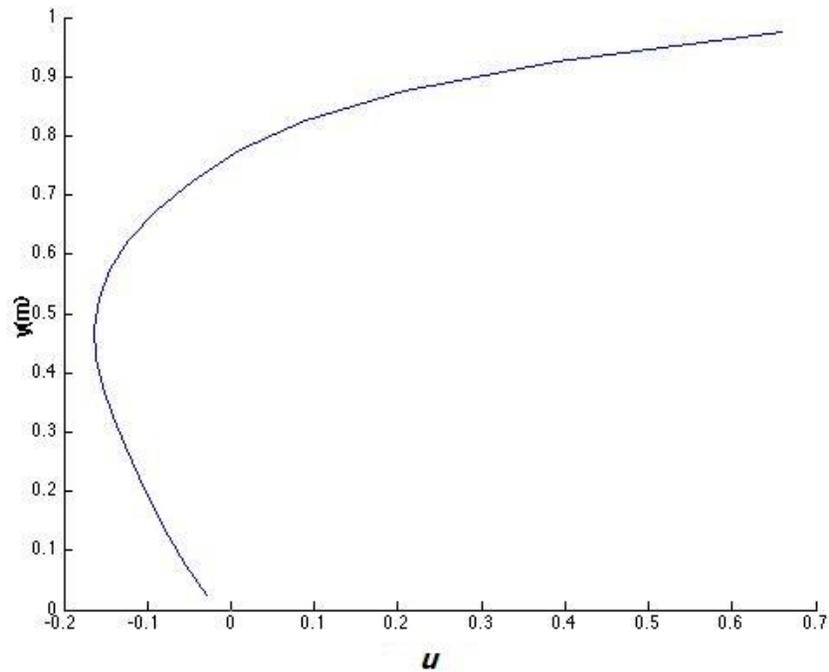

Figure 8 Velocity profile of the *x*-direction component (*u*) on the vertical centreline of the plane *z* = 0.5.

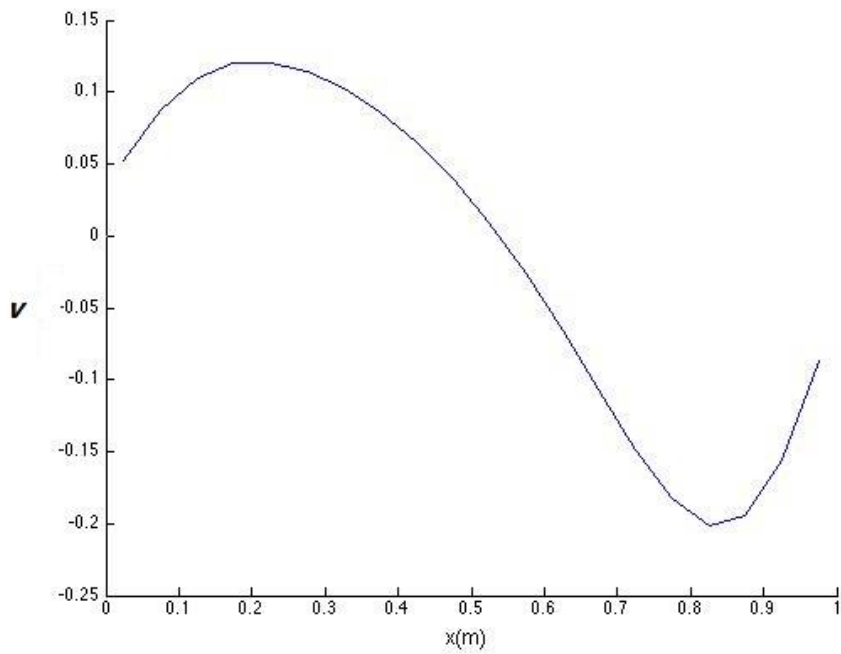

Figure 9 Velocity profile of the *y*-direction component (*v*) on the horizontal centreline of the plane *z* = 0.5.

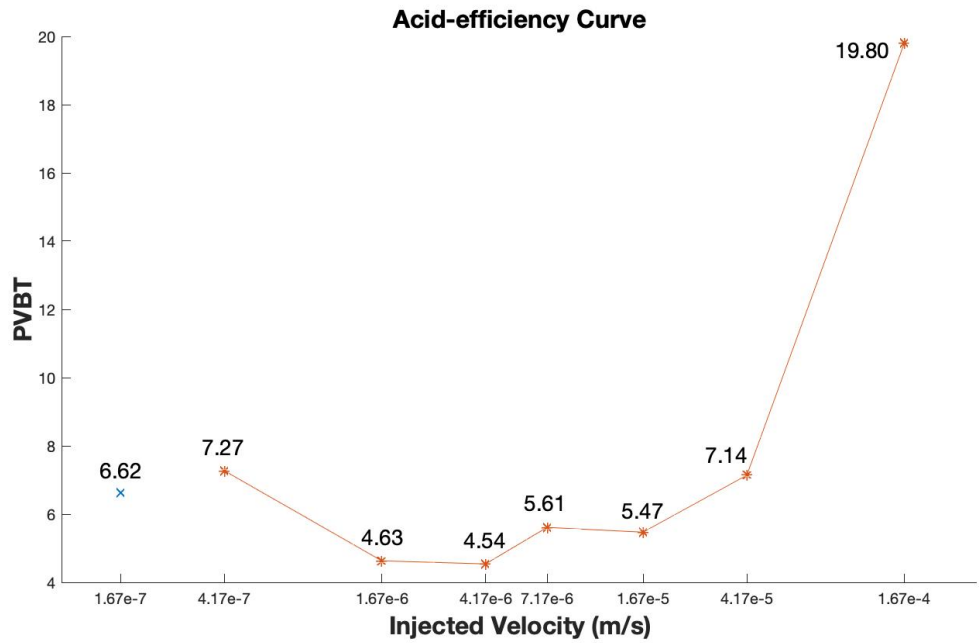

Figure 10 Acid-efficiency curve of the 2D linear flows. The numbers beside the points represent the values of PVBT.

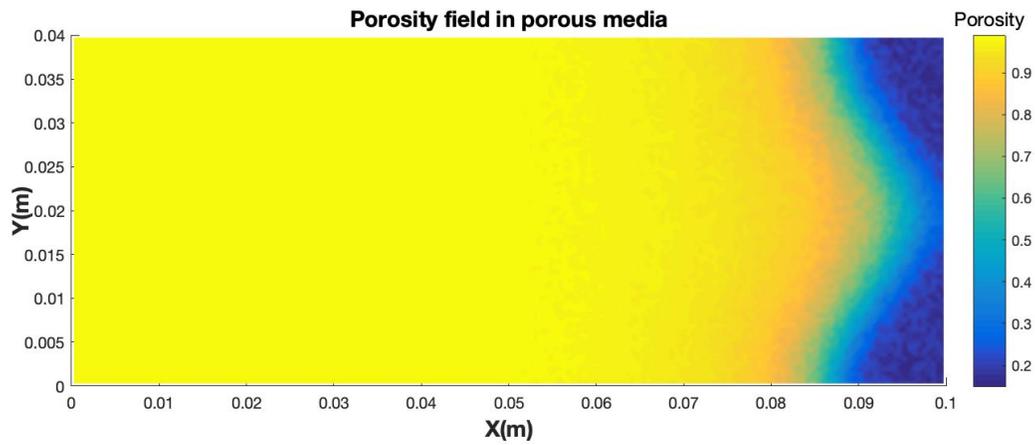

(a) $u_x = 4.17 \times 10^{-7}$ m/s, face dissolution.

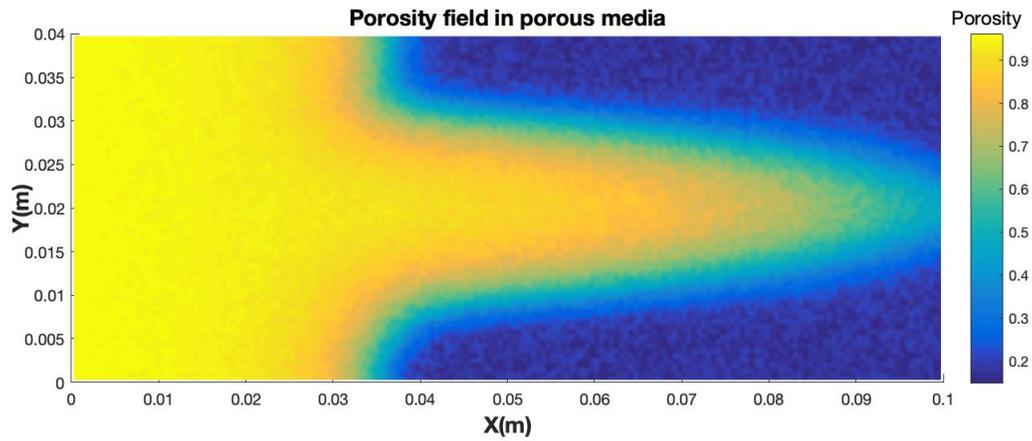

(b) $u_x = 1.67 \times 10^{-6}$ m/s, conical wormhole.

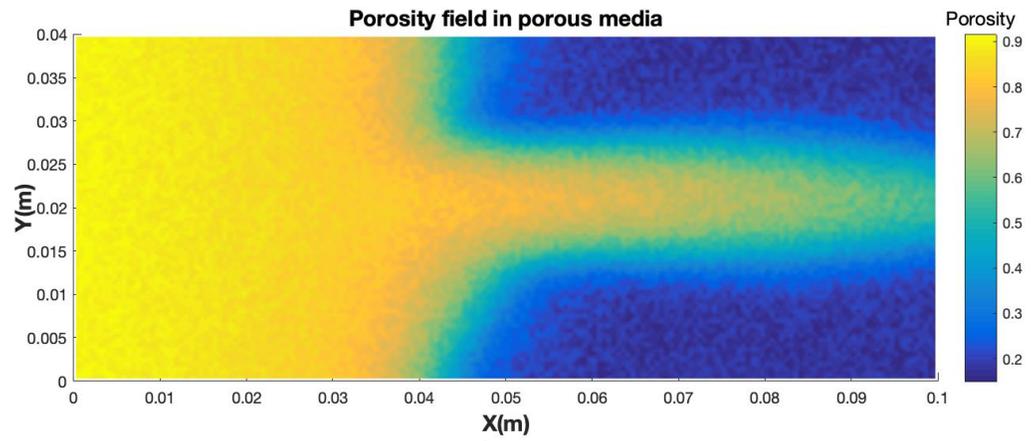

(c) $u_x = 4.17 \times 10^{-6}$ m/s, dominant wormhole.

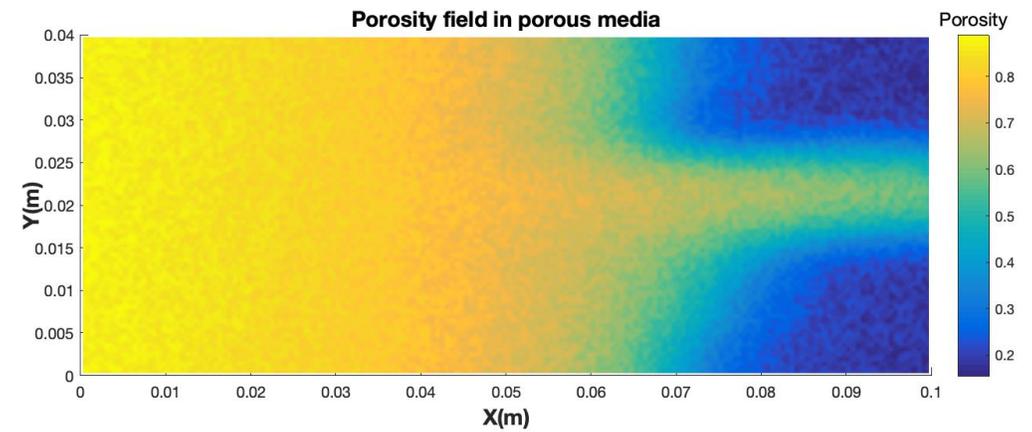

(d) $u_x = 7.17 \times 10^{-6}$ m/s, ramified wormhole.

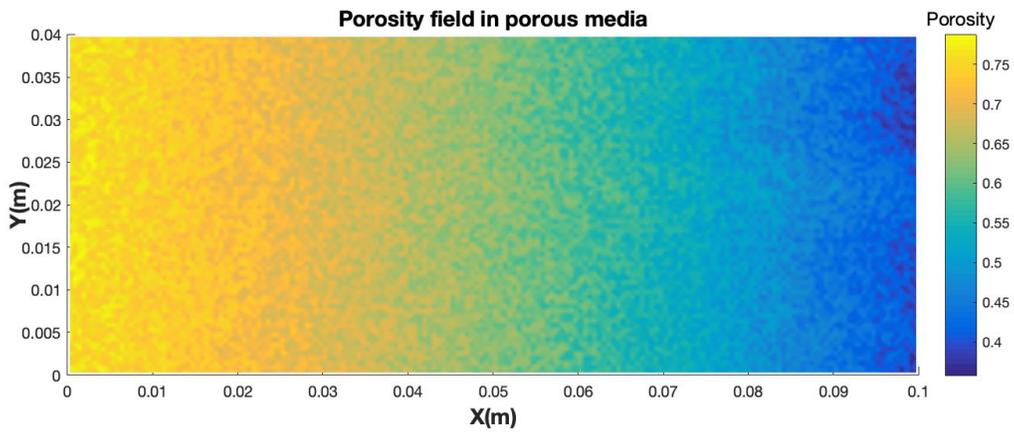

(e) $u_x = 1.67 \times 10^{-5}$ m/s, uniform dissolution.

Figure 11 Porosity profiles at breakthrough for five different injected velocities of the 2D linear flows.

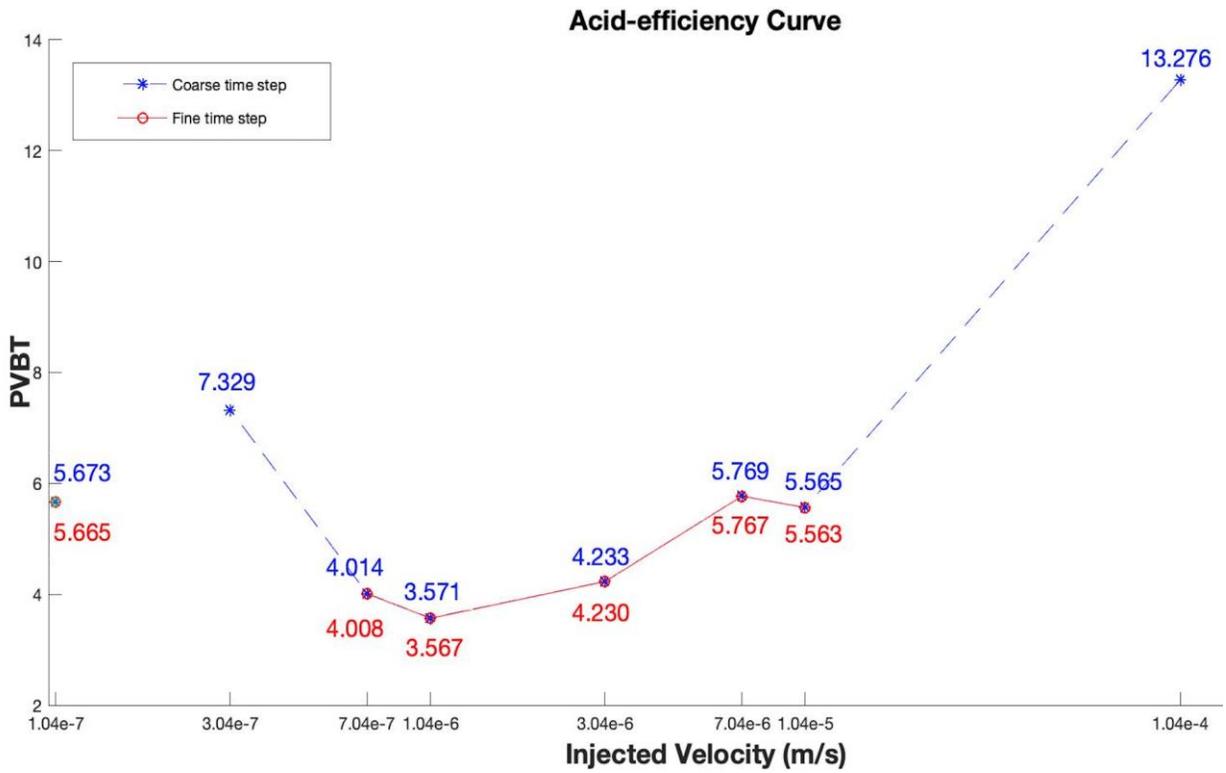

Figure 12 Acid-efficiency curves of the 3D linear flows. The numbers beside the points represent the values of PVBT.

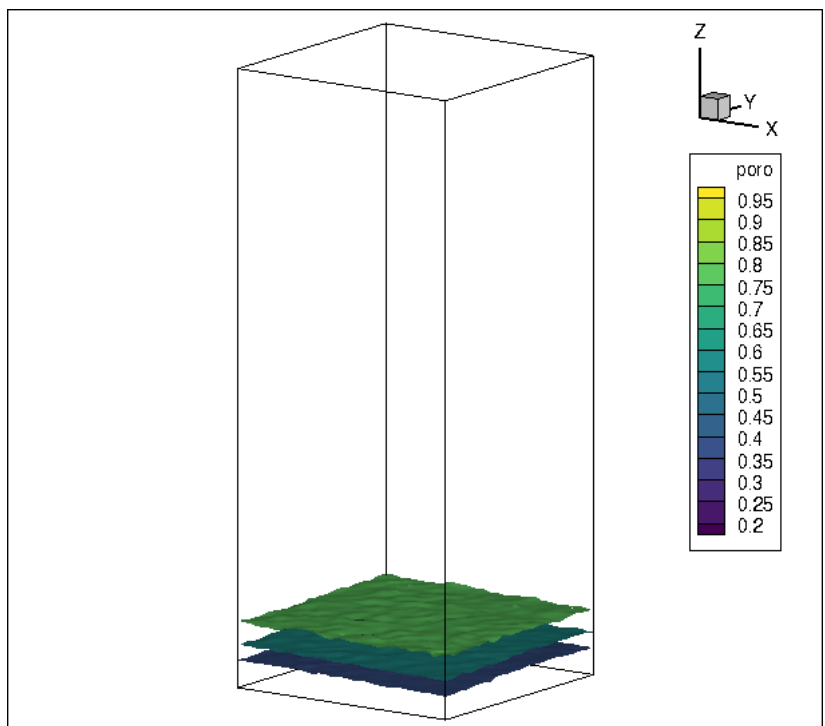

(a) $u_z = 3.04 \times 10^{-7}$ m/s, face dissolution.

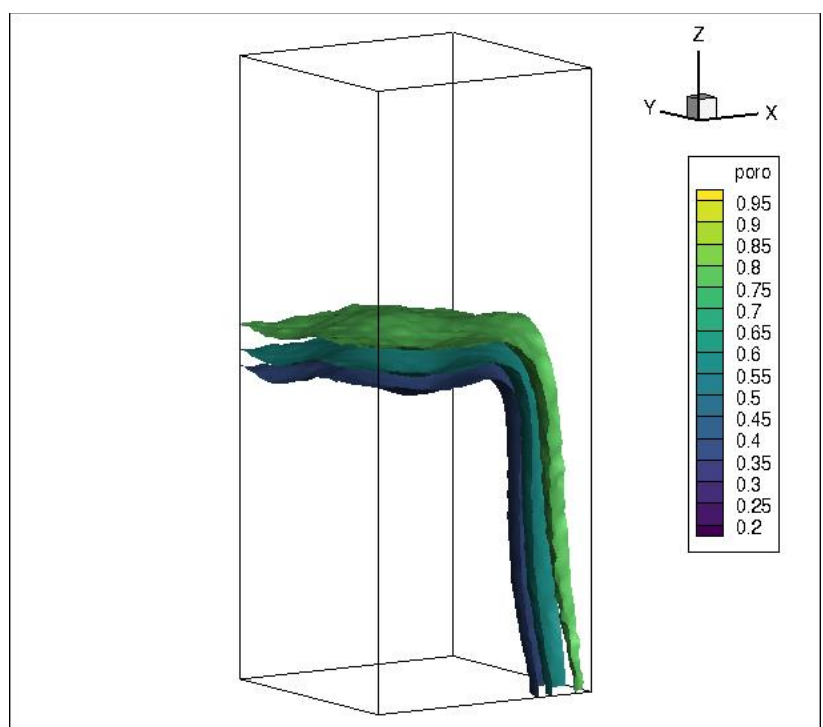

(b) $u_z = 7.04 \times 10^{-7}$ m/s, conical wormhole.

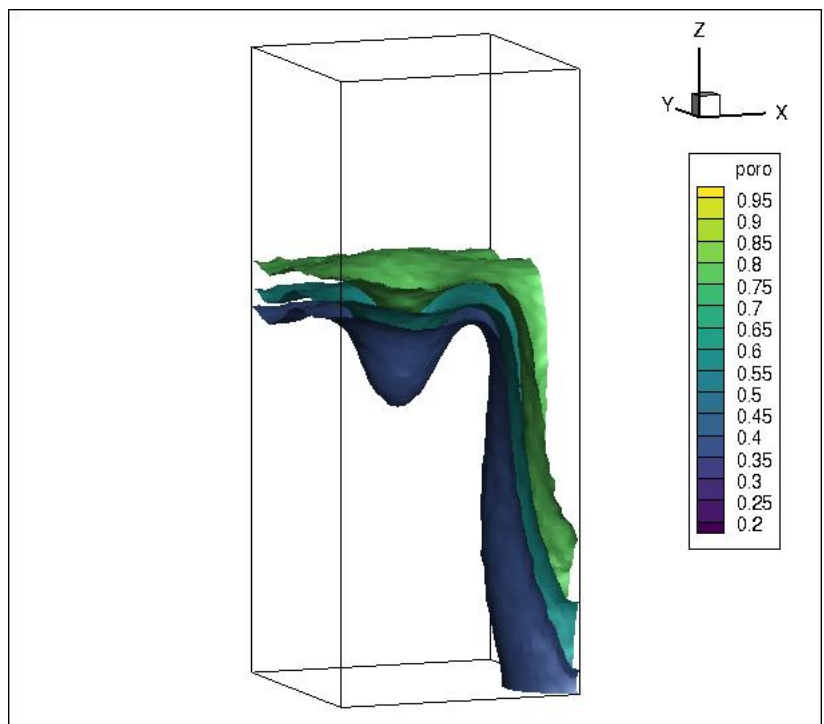

(c) $u_z = 1.04 \times 10^{-6}$ m/s, dominant wormhole.

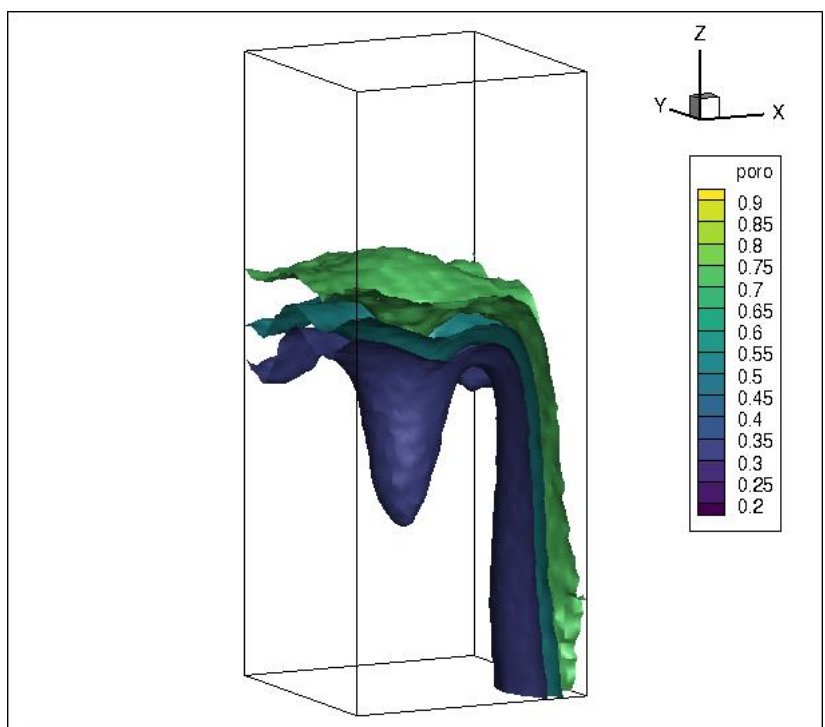

(d) $u_z = 3.04 \times 10^{-6}$ m/s, ramified wormhole.

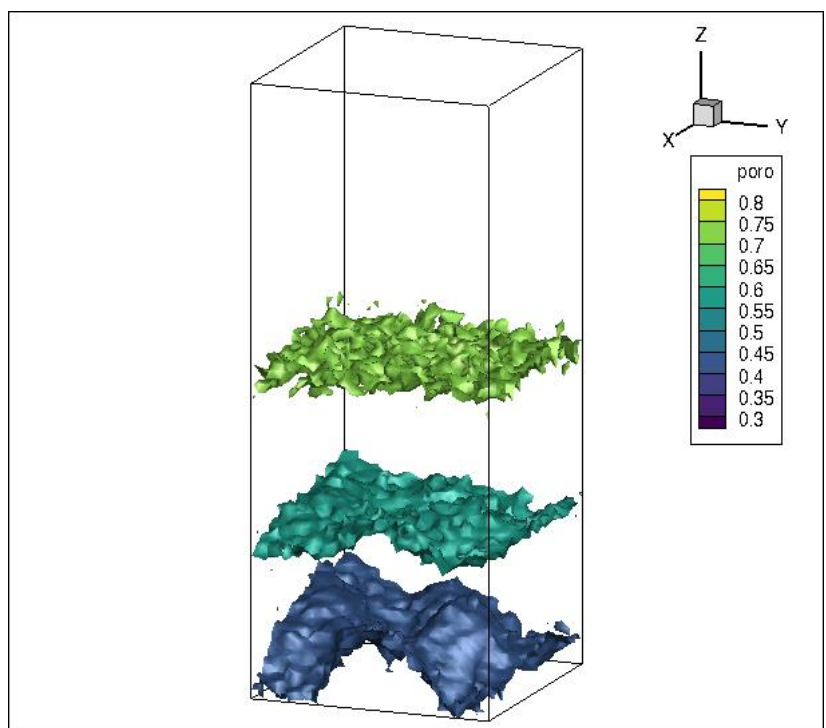

(e) $u_z = 1.04 \times 10^{-5}$ m/s, uniform dissolution.

Figure 13 Porosity iso-surfaces at breakthrough for five different injected velocities of the 3D linear flows.

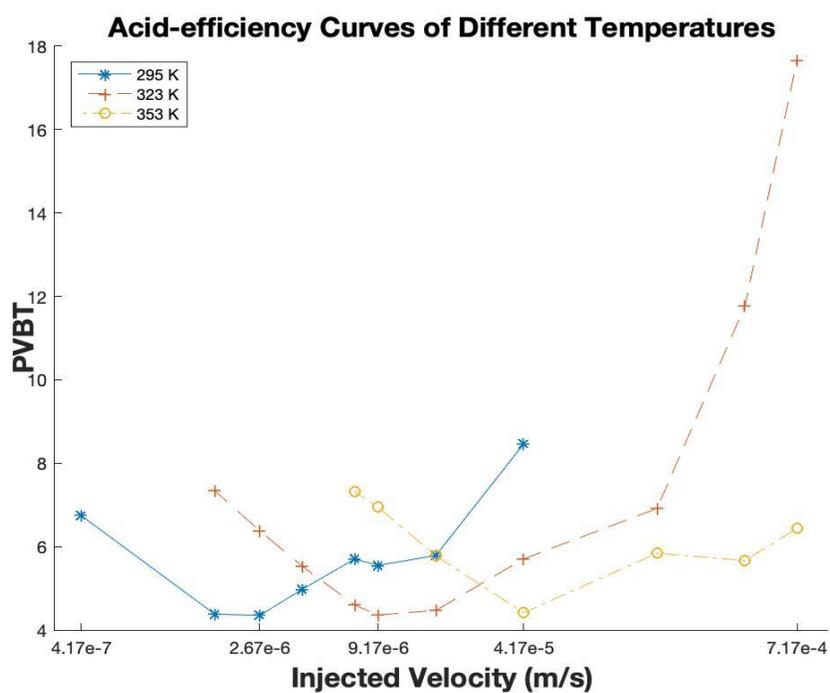

Figure 14 Acid-efficiency curves of different temperatures in isothermal conditions.

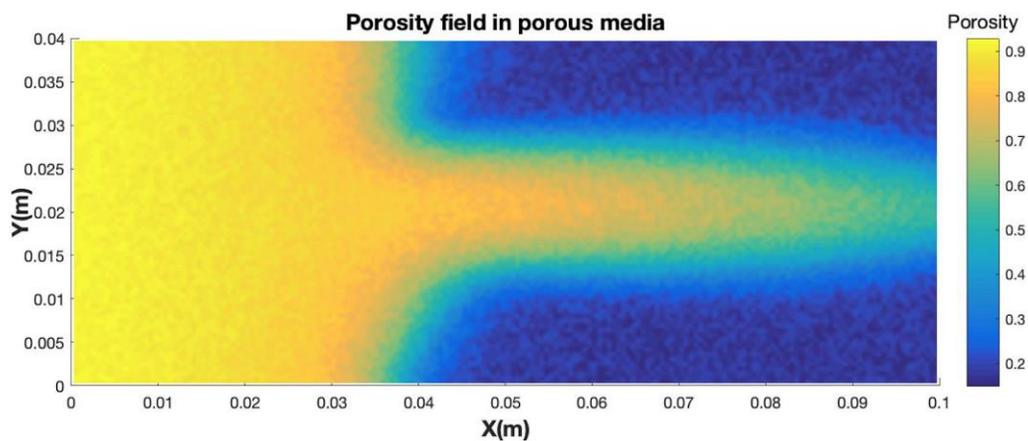

(a) 295 K, $u_x = 2.67 \times 10^{-6}$ m/s.

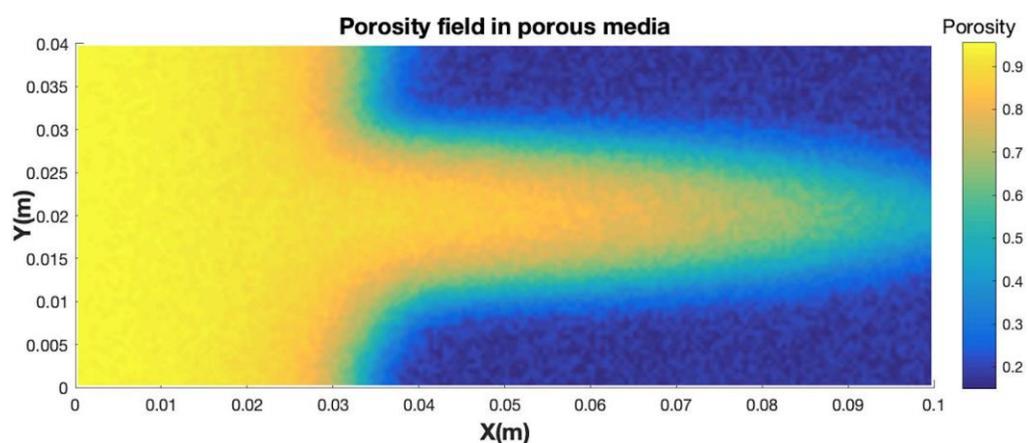

(b) 323 K, $u_x = 9.17 \times 10^{-6}$ m/s.

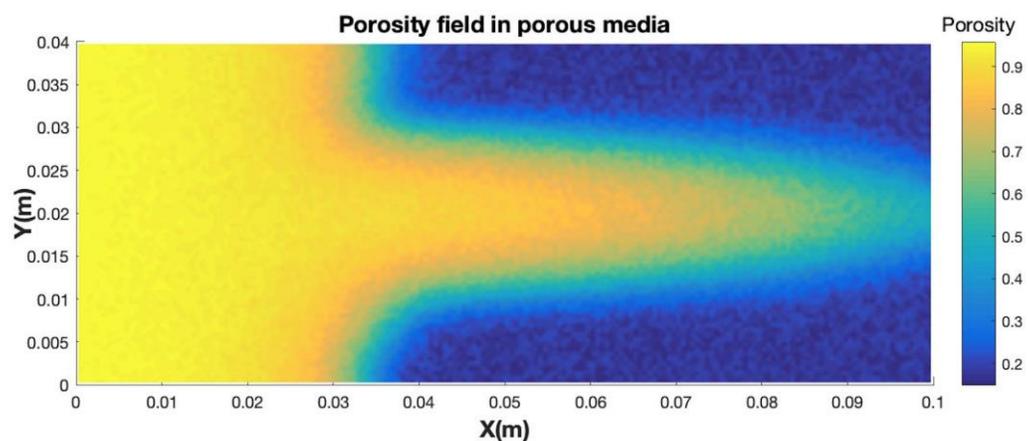

(c) 353 K, $u_x = 4.17 \times 10^{-5}$ m/s.

Figure 15 Porosity profiles at breakthrough in the optimal injected velocity for three different temperatures.

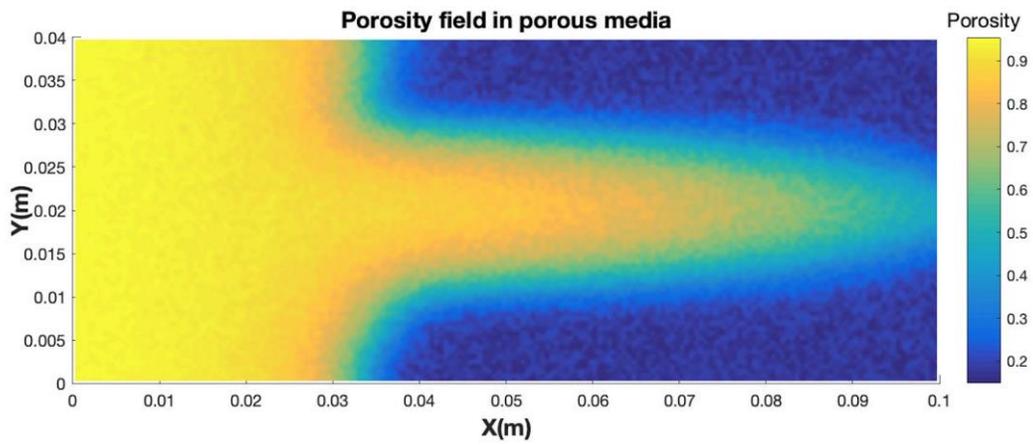

(a) 295 K

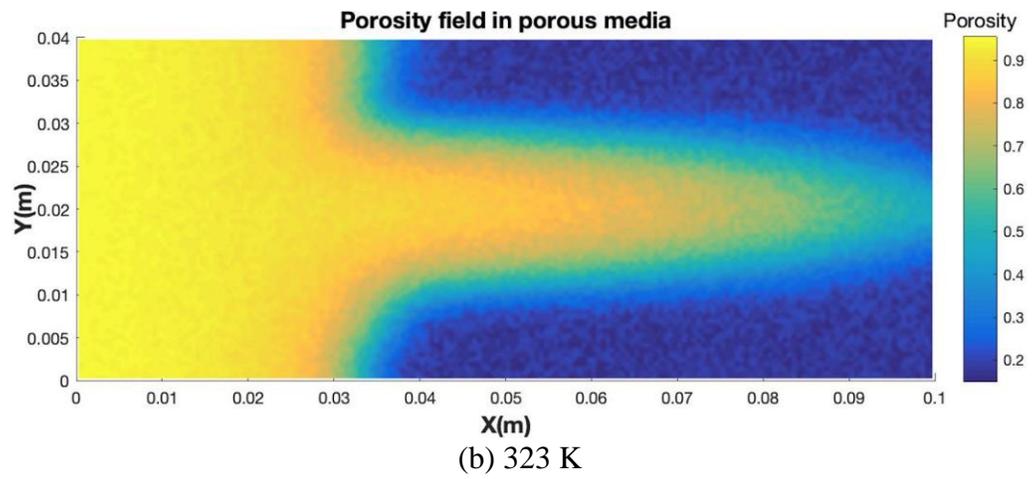

(b) 323 K

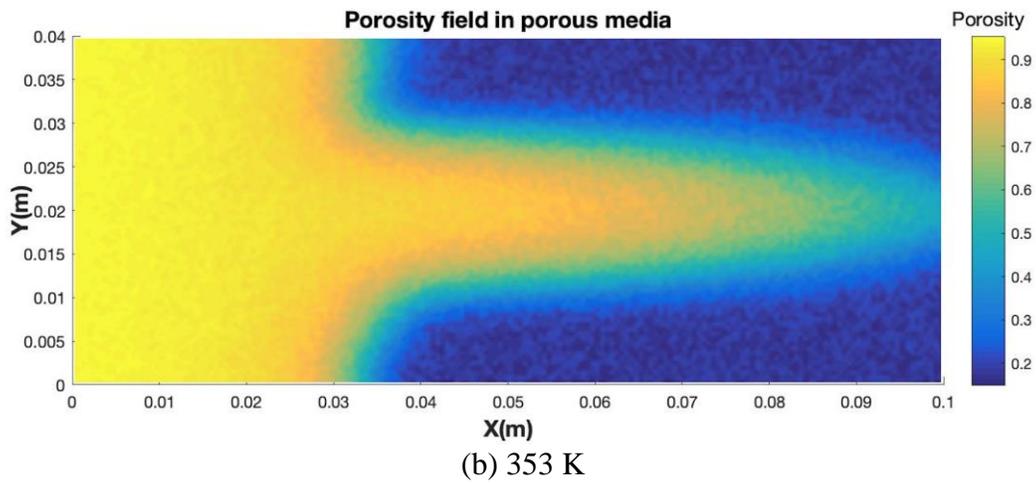

(b) 353 K

Figure 16 Porosity profiles at breakthrough for three different initial matrix temperatures. The injected acid temperature is 323 K, and its injected velocity is $9.17 \times 10^{-6}$ m/s.

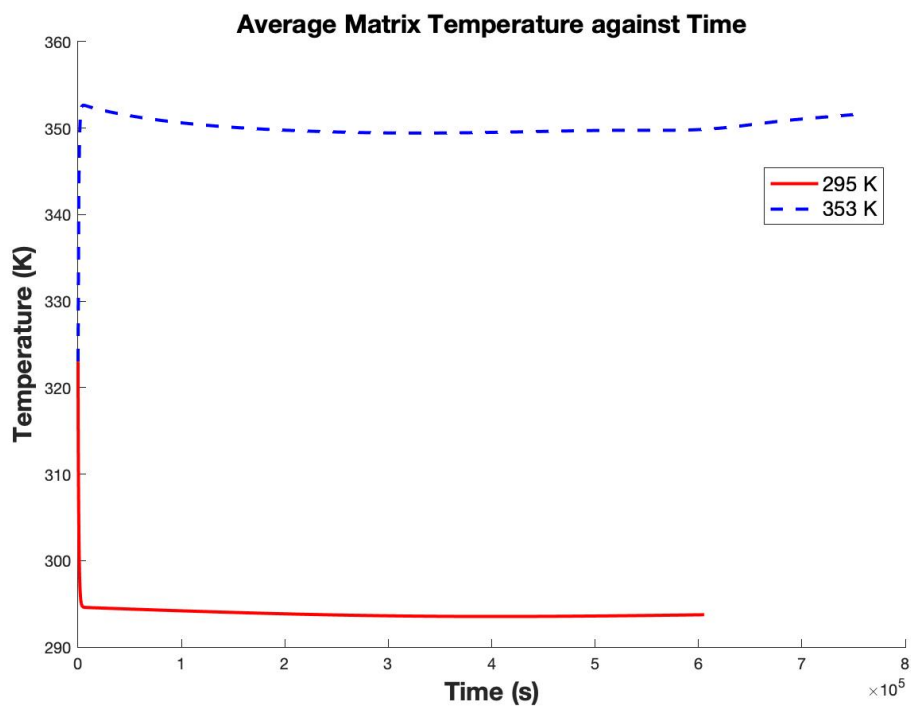

Figure 17 Average matrix temperature against time. The solid curve represents the case where the temperature of the injected acid is 295 K, and the dashed curve represents the case where the temperature of the injected acid is 353 K. The initial matrix temperature is 323 K.

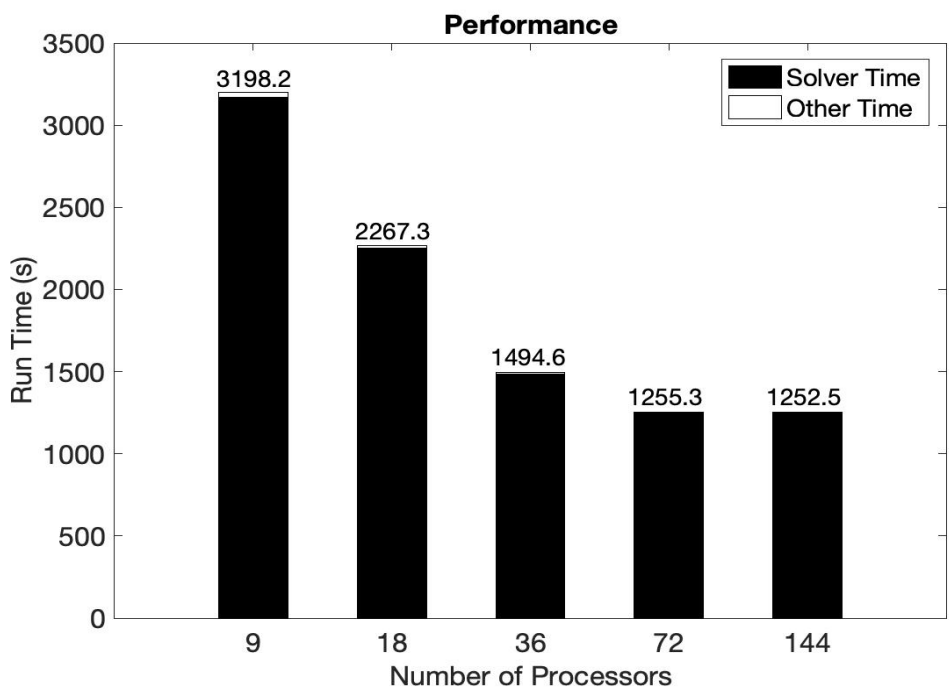

Figure 18 Performance of the 3D parallel code. The numbers above the bars represent the run time, which is the sum of the solver time and the other time.

# Tables

Table 1 Nomenclature

| Notation | Meaning |
| --- | --- |
| $p$ | pressure |
| $\mu$ | fluid viscosity |
| $K$ | permeability value |
| $\boldsymbol{u}$ | velocity vector |
| $Re$ | Reynolds number |
| $Da$ | Darcy number |
| $\phi$ | porosity |
| $\rho_f$ | mass density of the fluid |
| $F$ | Forchheimer coefficient |
| $t$ | time |
| $\boldsymbol{g}$ | gravity vector |
| $C_f$ | cup-mixing concentration of the acid |
| $\boldsymbol{D}_e$ | effective dispersion tensor |
| $d_m$ | molecular diffusion coefficient |
| $d_l$ | longitudinal dispersion coefficient |
| $d_t$ | transverse dispersion coefficient |
| $\alpha_{OS}$ | constant depending on pore connectivity |
| $\lambda_X$ | constant depending on the structure of the medium |
| $\lambda_T$ | constant depending on the structure of the medium |
| $r_p$ | pore radius |
| $\boldsymbol{E}$ | orthogonal projection along the velocity |
| $\boldsymbol{I}$ | identity matrix |
| $k_c$ | local mass-transfer coefficient |
| $a_v$ | interfacial surface area per unit volume |
| $C_s$ | concentration of the acid at the fluid-solid interface |
| $T$ | temperature |
| $R(C_s, T)$ | reaction rate |
| $\alpha$ | dissolving power of the acid |
| $\rho_s$ | mass density of the solid phase |
| $k_s$ | surface reaction rate |
| $Sh$ | Sherwood number |
| $Sh_\infty$ | asymptotic Sherwood number |
| $S_c$ | Schmidt number |
| $\tau$ | time step |
| $H_r(T)$ | reaction heat |
| $\lambda_f$ | heat conduction coefficient of the fluid phase |
| $\lambda_s$ | heat conduction coefficient of the solid phase |
| $\lambda$ | average heat conduction coefficient |
| $\theta_f$ | heat capacity of the fluid phase |
| $\theta_s$ | heat capacity of the solid phase |
| $\vartheta_f$ | amount of heat per unit volume of the fluid phase |
| $\vartheta_s$ | amount of heat per unit volume of the solid phase |
| $\vartheta$ | total amount of heat per unit volume |
| $E_g$ | activation energy |
| $R_g$ | molar gas constant |

Table 2 Experimental parameters

| Parameter | Value |
|---|---|
| $p_0$ | $1.52 \times 10^7$ Pa |
| $\mu$ | $1.0 \times 10^{-3}$ kg/(m·s) |
| $K_0$ | $9.869233 \times 10^{-16}$ m$^2$ |
| $\bar{\phi}_0$ | $1.8 \times 10^{-1}$ |
| $\rho_f$ | $1.01 \times 10^3$ kg/m$^3$ |
| $C_f$ | $5.0 \times 10^2$ mol/m$^3$ |
| $d_m$ | $3.6 \times 10^{-9}$ m$^2$/s |
| $\alpha_{OS}$ | $5.0 \times 10^{-1}$ |
| $\lambda_X$ | $5.0 \times 10^{-1}$ |
| $\lambda_T$ | $1.0 \times 10^{-1}$ |
| $r_0$ | $1.0 \times 10^{-6}$ m |
| $a_{v0}$ | $5 \times 10^{-1}$ m$^{-1}$ |
| $\alpha$ | $5.0 \times 10^{-2}$ kg/mol |
| $\rho_s$ | $2.71 \times 10^3$ kg/m$^3$ |
| $k_s$ | $2.0 \times 10^{-3}$ m/s |
| $Sh_\infty$ | 3.66 |
| $\lambda_f$ | $5.8 \times 10^{-1}$ W/(m · K) |
| $\lambda_s$ | 5.526 W/(m · K) |
| $\theta_f$ | $4.184 \times 10^3$ J/(kg · K) |
| $\theta_s$ | $2.0 \times 10^2$ J/(kg · K) |
| $k_{s0}$ | $2.0 \times 10^{-3}$ m/s (298 K) |
| $d_{m0}$ | $3.6 \times 10^{-9}$ m$^2$/s (298 K) |
| $E_g$ | $5.02416 \times 10^4$ J/mol |
| $R_g$ | 8.314 J/(K · mol) |

Table 3 Time steps for the injected velocities of the 2D linear flows

| Velocity (m/s) | Time step (s) |
|---|---|
| $1.67 \times 10^{-7}$ | 1579 |
| $4.17 \times 10^{-7}$ | 643 |
| $1.67 \times 10^{-6}$ | 150 |
| $4.17 \times 10^{-6}$ | 60 |
| $7.17 \times 10^{-6}$ | 16 |
| $1.67 \times 10^{-5}$ | 16.5 |
| $4.17 \times 10^{-5}$ | 6.43 |
| $1.67 \times 10^{-4}$ | 1.5 |

Table 4 Time steps for the injected velocities of the 3D linear flows

| Velocity (m/s) | Coarse time step (s) | Fine time step (s) |
|---|---|---|
| $1.04 \times 10^{-7}$ | 9322 | 4661 |
| $3.04 \times 10^{-7}$ | 3611 | / |
| $7.04 \times 10^{-7}$ | 984 | 492 |
| $1.04 \times 10^{-6}$ | 593.2 | 296.6 |
| $3.04 \times 10^{-6}$ | 237.2 | 118.6 |
| $7.04 \times 10^{-6}$ | 139 | 69.5 |
| $1.04 \times 10^{-5}$ | 91.6 | 45.8 |
| $1.04 \times 10^{-4}$ | 21.2 | / |

Table 5 Values of PVBT. The first row represents injected velocities, and their unit is m/s.

| | $4.17 \times 10^{-7}$ | $1.67 \times 10^{-6}$ | $2.67 \times 10^{-6}$ | $4.17 \times 10^{-6}$ | $7.17 \times 10^{-6}$ | $9.17 \times 10^{-6}$ | $1.67 \times 10^{-5}$ | $4.17 \times 10^{-5}$ | $1.67 \times 10^{-4}$ | $4.17 \times 10^{-4}$ | $7.17 \times 10^{-4}$ |
|---|---|---|---|---|---|---|---|---|---|---|---|
| 295 K coarse-time step | 6.750 | 4.387 | 4.352 | 4.982 | 5.703 | 5.553 | 5.789 | 8.465 | / | / | / |
| 295 K fine-time step | 6.745 | 4.385 | 4.350 | 4.981 | 5.703 | 5.552 | 5.788 | 8.465 | / | / | / |
| 323 K coarse-time step | / | 7.348 | 6.387 | 5.530 | 4.602 | 4.364 | 4.473 | 5.708 | 6.917 | 11.771 | 17.666 |
| 323 K fine-time step | / | 7.344 | 6.383 | 5.528 | 4.601 | 4.362 | 4.471 | 5.708 | 6.916 | 11.771 | 17.665 |
| 353 K coarse-time step | / | / | / | / | 7.321 | 6.952 | 5.785 | 4.418 | 5.844 | 5.669 | 6.438 |
| 353 K fine-time step | / | / | / | / | 7.319 | 6.947 | 5.783 | 4.416 | 5.843 | 5.668 | 6.437 |

Table 6 Values of PVBT for non-isothermal conditions

| Injected acid temperature (K) | Initial matrix temperature (K) | PVBT |
|---|---|---|
| 295 | 323 | 5.552 |
| 353 | 323 | 6.948 |
| 323 | 295 | 4.362 |
| 323 | 353 | 4.362 |

# Acknowledgement

For computer time, this research used the resources of the Supercomputing Laboratory at King Abdullah University of Science & Technology (KAUST) in Thuwal, Saudi Arabia.